\documentclass[traditabstract]{aa} 
\pdfoutput=1
\usepackage{graphicx}
\usepackage{longtable}
\usepackage{txfonts}

\usepackage{booktabs}
\begin{document}
   \title{HST/ACS color-magnitude diagrams of candidate intermediate-age M 31 globular clusters.
       \thanks{Based on observations made with the NASA/ESA {\em Hubble Space Telescope}, obtained 
            from the data archive at the Space Telescope Science Institute. STScI is operated
            by the Association of Universities for Research in Astronomy, Inc., under NASA 
            contract NAS 5-26555. These observations are associated with program GO-10631 
            [P.I.: T. Puzia].}
	    \fnmsep\thanks{Photometric catalogs are available in electronic form
	     at the CDS via anonymous ftp to cdsarc.u-strasbg.fr (130.79.128.5)
             or via http://cdsweb.u-strasbg.fr/cgi-bin/qcat?J/A+A/ and 
	     at http://www.bo.astro.it/M31/hstcatalog/}}
   \subtitle{The role of blue horizontal branches}

   \author{S. Perina\inst{1}, S. Galleti\inst{1}, F. Fusi Pecci\inst{1}, 
M. Bellazzini\inst{1}, L. Federici\inst{1}, \and A. Buzzoni\inst{1}
          }
	  
   \offprints{sibilla.perina2@unibo.it}

   \institute{INAF - Osservatorio Astronomico di Bologna,
              Via Ranzani 1, 40127 Bologna, Italy\\
          \email{sibilla.perina2@unibo.it}
      }

     \authorrunning{Perina et al.}
  \titlerunning{HST/ACS CMDs of candidate intermediate-age M31 globular clusters.}

   \date{Submitted to A\&A }

\abstract{We present deep ($V\simeq 28.0$) BV photometry obtained with the wide field channel of the 
Advanced Camera for Surveys on board HST for four M31 globular clusters that were identified 
as candidate intermediate-age (age$\sim 1-9$~Gyr) by various authors, based on their integrated 
spectra and/or broad/intermediate-band colors. 
Two of them (B292 and B350) display an obvious blue horizontal branch, indicating that they are as 
old as the oldest Galactic globulars. 
On the other hand, for the other two (B058 and B337), which display red horizontal branches, it was not 
possible either to confirm or disconfirm the age estimate from integrated spectra.
The analysis of the distribution in the spectral indices Mg2 and H$\beta$ of the M31 and Milky Way 
clusters whose horizontal branch can be classified as red or blue based on existing CMDs, strongly 
suggests that classical age diagnostics from integrated spectra may be significantly influenced by 
the HB morphology of the clusters and can lead to erroneous age-classifications.
We also provide the CMD for another two clusters that fall into the field of the main targets, B336, an 
old and metal-poor globular with a significant population of RR~Lyrae variables, and the newly 
discovered B531, a cluster with a very red red giant branch.}


   \maketitle

\section{Introduction} 
\label{int}

Nearly all types of galaxies contain globular clusters (GC), from dwarfs to
giants and from the earliest to the latest types. Our most in-depth
understanding of individual GCs comes from studies of the Milky Way
(MW; see, for example, Dotter et al.~\cite{dotter} and references therein).

Among the external galaxies, M31 plays a special role in our studying of GCs. 
It is our nearest bright spiral galaxy neighbor, it is the most prominent member in the Local
Group (LG), and it hosts the largest population of GCs: 544 confirmed GCs are listed in the latest version of the Revised Bologna Catalog\footnote{\tt http://www.bo.astro.it/M31/} (RBC v4.0, Galleti et al.~\cite{rbc}). 
At the distance of M31 ($\sim783$ kpc, from McConnachie et al.~\cite{mcc05}), GCs have almost stellar appearance 
(10 pc correspond to $\sim 2.6 \arcsec$), hence the diagnostics based on integrated light that are used for more distant galaxies can be homogeneously applied here. On the other hand, M31 is also close enough that individual stars in GCs can be 
resolved and measured with the Hubble Space Telescope (HST). 
In this framework, the GCs system of
M31 is a fundamental test bed for checking the consistency between the cluster parameters (in particular age and metallicity) derived from the integrated light and those obtained from the analysis of the color-magnitude diagram (CMD) of individual cluster stars.

The most powerful method to determine cluster ages relies on the accurate
location of the main-sequence turn-off (MSTO) point (Renzini \& Fusi Pecci \cite{rfp}).
Unfortunately, this method is limited to the nearest GCs where individual stars
can be resolved and measured down to at least $\simeq$1-2 magnitudes fainter than the MSTO.
At the distance of M31 the task is made very challenging owing to the effects
of crowding and the intrinsic faintness of the feature. At present only one of 
M31's GC has a CMD down to the TO, which has been achieved only with
$\sim$3.5 days of HST/ACS integration time (B379; Brown et al.~\cite{brown04})\footnote{It is interesting to note that B379 was found by Brown et al.(~\cite{brown04}) to be 2-3~Gyr younger than Galactic GCs of the same metallicity, with an absolute age of $10^{+2.5}_{-1}$~Gyr. Re-analyzing the same data and using different stellar models, Ma et al.~(\cite{ma10}) estimated an age of $11.0\pm 1.5$~Gyr.}.

In general, typical CMDs of M31 GCs obtained by HST data barely reach 
the base of the red giant branch (RGB), including most horizontal branch (HB) stars, except the bluest/faintest
ones, when present (Ajhar et al.~\cite{ajh96}; Rich et al.~\cite{ric96a};
Fusi Pecci et al.~\cite{fp96}; Holland et al.~\cite{holl97}; 
Jablonka et al.~\cite{jab00}; Meylan et al.~\cite{mey01}; Rich et al.~\cite{rich05};
Perina et al.~\cite{P09b}; Huxor et al.~\cite{hux04,hux05,hux08}; 
Galleti et al.~\cite{b514}; Mackey et al.~\cite{mack06,mack07}).
In this magnitude range the HB morphology can be used as a rough age indicator, because the presence of RR~Lyrae variables and/or blue HB stars implies low-mass progenitors, hence old ages.
In a recent analysis, Dotter et al.(~\cite{dotter}) conclude that all Galactic GCs with a blue horizontal branch have ages $\ga 12.0$~Gyr, on a scale in which the oldest GC is $\simeq 13.5$~gyr old.

To date, there are relatively few M31 GCs with a CMD (44).
As a consequence, our overall characterization of the GC system of that galaxy mainly relies on 
the analysis of integrated colors and spectra
(see Galleti et al.~\cite{rbc}; Puzia et al.~\cite{puzia}; Caldwell et al.~\cite{C09},~\cite{C11}, and references therein). 
A widely used technique is based on Lick spectral indices (Burstein et al.~\cite{lick1}; Faber et al.~\cite{lick2}; Worthey~\cite{worthey}). Typically,
indices that are mainly sensitive to age ($H_{\beta}$, $H_{\gamma}$) are used in combination with
others that are more sensitive to metallicity (Mgb, [MgFe]), attempting to
break the well-known age-metallicity degeneracy. At fixed metallicity, 
stronger Balmer lines indicate a hotter MSTO, hence younger ages 
(see for example Puzia et al.~\cite{puzia} and Caldwell et al.~\cite{C09}, \cite{C11}).
Another widely used method to age-date extragalactic star clusters relies on multi-band photometry: the age can be derived from the spectral energy distributions (SED)
measured in broad-band photometric systems, by comparing them with simple
stellar population (SSP\footnote{A simple stellar population is an ideal population of stars that have the same age and chemical composition (Renzini \& Fusi Pecci \cite{rfp}).}) synthesis models (e.g. Jiang et al.~\cite{jiang03}; Fan et al.~\cite{fan06,fan08,fan10}; 
Ma et al.~\cite{ma07,ma09}; and Wang et al.~\cite{wang10}). Clearly, photometry-based technique are prone to errors owing to the uncertainty in the interstellar extinction, while spectral indices are virtually free from this effect.

In a recent analysis adopting the most up-to-date theoretical tools and methodology, 
Puzia et al.~(\cite{puzia}; P05 hereafter) derived spectroscopic ages, metallicities and $[\alpha/Fe]$ ratios  for 70 GCs in M31 based on Lick indices. Within this sample, the authors find a population  of $\sim$20 GCs 
with ages between $\sim 5$ and 9 Gyr and mean metallicity of $[Z/H] \sim -0.6$. 
Independently, Burstein et al.~(\cite{burst04}) and Beasley et al.~(\cite{beas05}) also claimed to have found two
(B232 and B311) and six (B126, B292, B301, B337, NB16, NB67) {\em candidate} intermediate-age GCs in M31\footnote{It is interesting to note that Beasley et al.~(\cite{beas05}) find remarkably different age estimates from the same data, depending on the set of theoretical models they compare with. In particular, ages obtained by comparison with the Bruzual \& Charlot~(\cite{bc03}) models are systematically lower than those derived from the comparison with Thomas et al.~(\cite{thomas03}) ones (see Table~\ref{tab:age}, below).}, respectively, in common with Puzia's sample. The presence of a population of intermediate-age GCs in M31 was also supported by other spectrophotometric analyses (Fan et al.~\cite{fan10}; Wang et al.~\cite{wang10} and reference therein; see Table~\ref{tab:age}). 

On the other hand, Strader et al.~(\cite{strad09}) showed that three of the candidate
intermediate-age M31 GCs studied by P05, Burstein et al.~(\cite{burst04}), and 
Beasley et al.~(\cite{beas05}) have $M/L_V$ and $V-K$ colors typical of old GCs. 
Twelve of the spectroscopically identified intermediate-age GC candidates were also detected by
Rey et al.~(\cite{rey07}) in far-ultraviolet (FUV) and/or near-ultraviolet (NUV) GALEX images. 
From the comparison of their UV photometry with theoretical models these authors concluded that most of these spectroscopically identified intermediate-age clusters may in fact be classical old GCs with many blue HB stars. These hot stars  would contribute to enhance the
UV flux as well as the strength of Balmer lines, thus mimicking younger ages. 
A possible example is provided by the cluster B311, reported to be 5 Gyr old by Burstein et al.~(\cite{burst04}), that reveals an extended HB in the CMD presented by Rich et al.~(\cite{rich05}), clearly not consistent with such a young age.
Therefore, deep HST imaging of intermediate-age GCs candidates is crucial to 
check if they are genuinely intermediate-age or, instead, old clusters with blue HB.
The presence of bright intermediate-age clusters in M31, if confirmed, would be especially interesting because these objects are not observed in our own Galaxy, possibly implying a significant difference in the star- (or, at least, cluster) formation history between the two galaxies.


   \begin{figure}
   \centering
   \includegraphics[width=9cm]{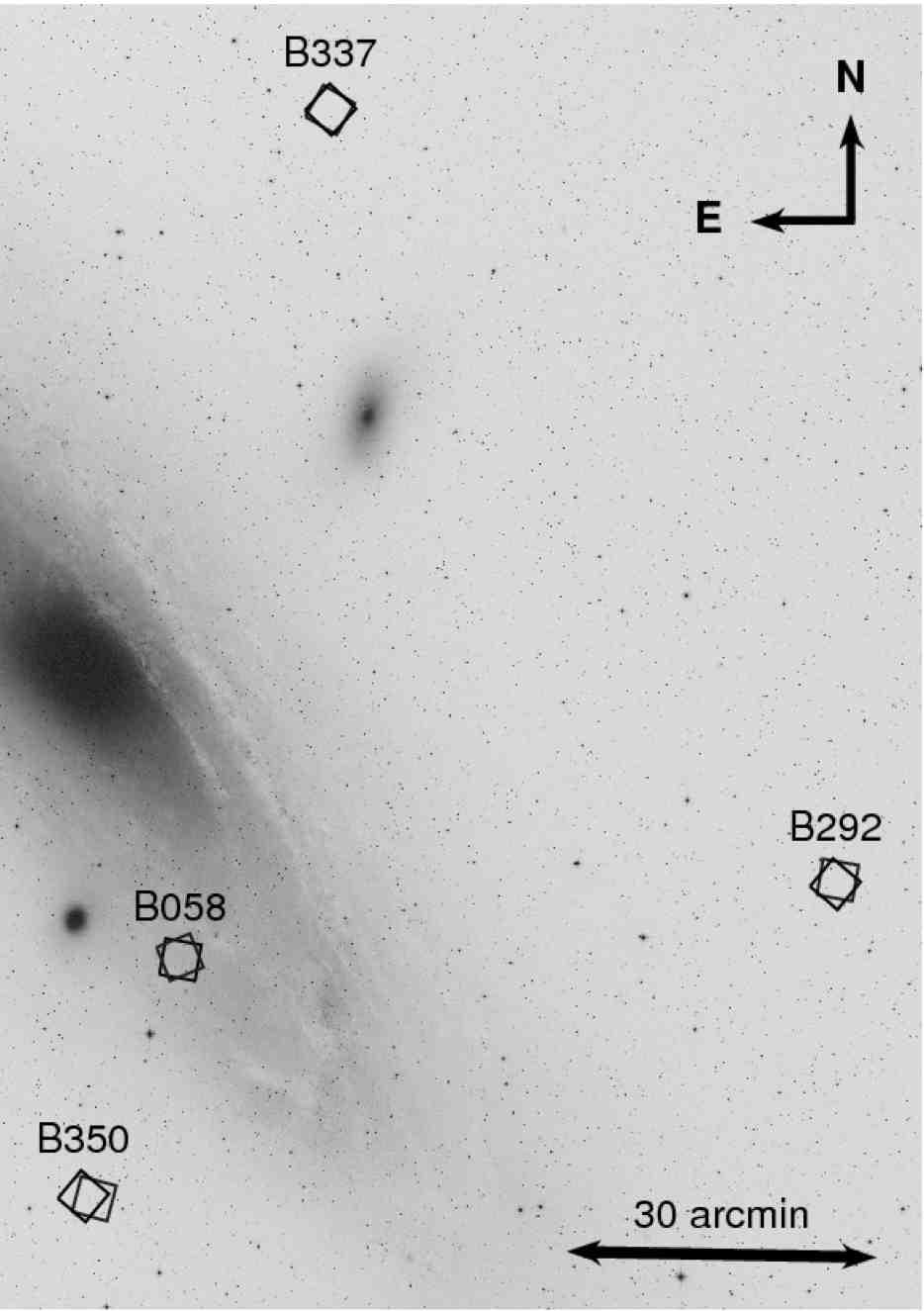}
      \caption{Location of the 4 target fields (open squares with the size of an ACS-WFC field) 
      projected against the body of M31. Note the different orientations of 
      the F606W and F435W exposures.}
         \label{fig:map}
   \end{figure}

In HST cycle 14 deep ACS/WFC imaging was acquired of four M31 GCs, which were identified by P05 as candidate intermediate-age cluster. 
In this paper we present the CMDs of the four main targets of this observational program, discussing the compatibility with the parameters obtained by P05 from the integrated spectroscopy (see Tables~\ref{tab:age} 
and ~\ref{tab:met}). 
We also present the CMDs for two other GCs that were in the target fields in both passbands: B336 and the newly detected cluster B531. Finally, we provide firmer classification for three additional objects listed in the RBC that were imaged only in one passband.

In Sect.~\ref{data} we describe the HST/ACS data, the adopted reduction procedure, the photometric uncertainties
and the completeness of the data.
Sect.~\ref{cmd} is devoted to describe the CMDs of the individual clusters, 
the field-decontamination procedure, and the method we used to estimate the metallicity, reddening, and distance.
In Sect.~\ref{RR-Lyrae} we describe the method for searching variable stars and the results for the individual clusters.  
Finally, Sect.~\ref{discu} is dedicated to the discussion of the results and conclusions.


\begin{table} 
 \centering
 \caption{Fundamental parameters of the four main targets of the survey.}
\label{tab:targets}

\begin{tabular} {l c c  c c} 
\hline
\\
ID  & RA (J2000) & Dec (J2000) &   V$_{int}$ & E(B-V)$^\ast$ \\
    &                       &        &   &               \\
\hline
\\
  B058-G119 & 00 41 53.01 & +40 47 08.6   & 14.97 & 0.13$\pm$0.01$^a$  \\  
  B292-G010 & 00 36 16.59 & +40 58 26.6   & 16.99 & 0.13$^b$    \\  
  B337-G068 & 00 40 48.45 & +42 12 11.5   & 16.73 & 0.06$\pm$0.02	    \\ 
  B350-G162 & 00 42 28.33 & +40 24 50.2   & 16.74 & 0.10$\pm$0.02	    \\  
\\
 \hline
\end{tabular} 
{\begin{flushleft}{The integrated apparent magnitudes of the clusters (V$_{int}$) are taken from RBC V4.0 (Galleti et al.\cite{rbc}).}
{~~$^{\ast}$ from Fan et al.~(\cite{fan08}).}
{~~$^a$ Rich et al.~(\cite{rich05}) find E(B-V)=0.12 for this cluster.}
{~~$^b$ from Caldwell et al.~(\cite{C11}). These authors provided reddening estimates also for the other clusters, in particular E(B-V)= 0.15, 0.10, and 0.12, for B058, B337, and B350, respectively (uncertainties not reported).}
\end{flushleft}}
\end{table}


   \begin{figure}
   \centering
   \includegraphics[width=9cm]{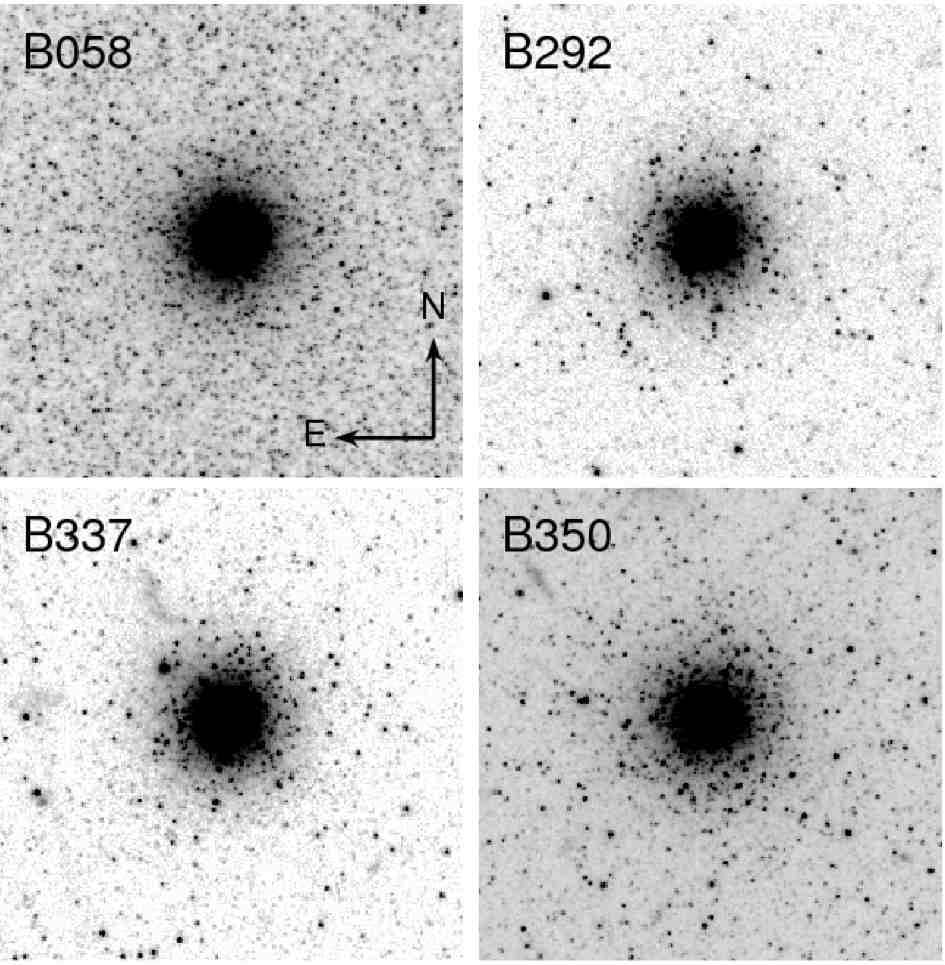}
      \caption{F606W images of the four candidate intermediate-age clusters that were the
      main target of the observations.
      The size of each image is 10${\arcsec}\times$~10${\arcsec}$.
      10${\arcsec}$ corresponds to 38 pc at the distance of M31 ($D=783$~kpc). 
      }
         \label{fig:targets}
   \end{figure}

   \begin{figure}
   \centering
   \includegraphics[width=9cm]{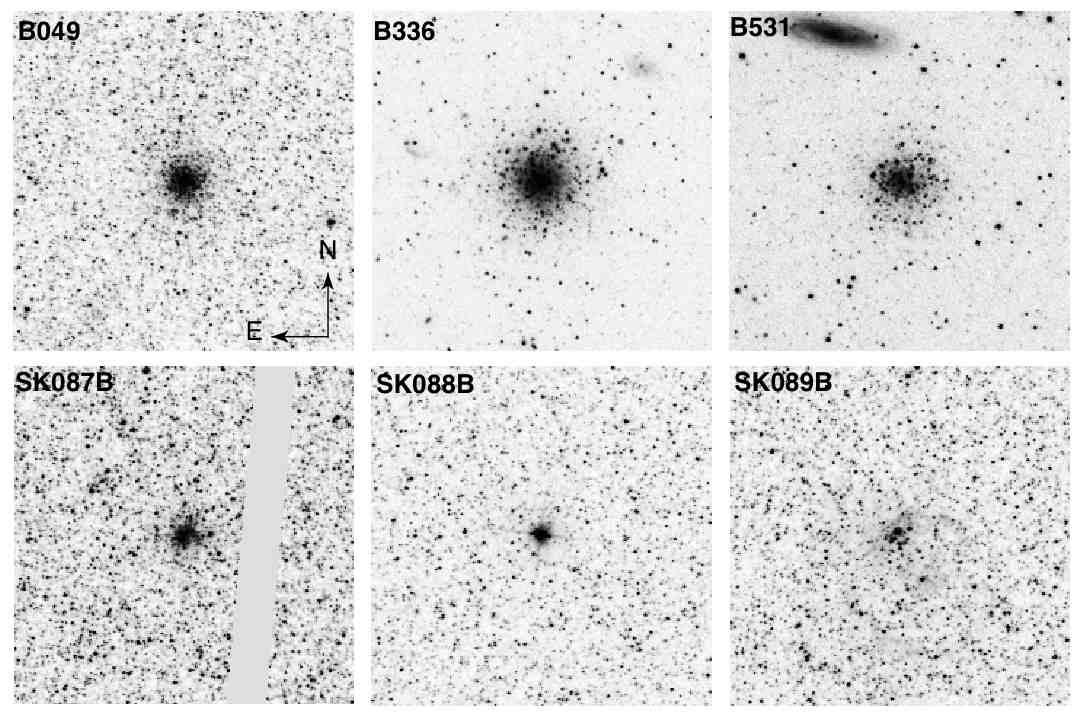}
      \caption{Stamp-images of other objects listed in the RBC that were included in the 
      analyzed WFC images 
      (B049, B336, SK087B, SK088B, SK089B) plus the newly discovered cluster B531.
      The size of each image is 10${\arcsec}\times$~10${\arcsec}$.
      }
         \label{fig:others}
   \end{figure}

\section{Observations and data reduction}  
\label{data}

The four main targets are the clusters B058, B292, B337, and B350. They were selected from Table~A.3 of  P05 as candidate intermediate-age GCs.The observations were performed under  program  GO-10631  (P.I.: Thomas Puzia),    
using  the  ACS  on  board  HST,  and  the  filters  F435W  (similar to Johnson B; four images with a total exposition time of  4800 sec for B058, and 4900 sec for the other three targets) 
and  F606W  (similar to Johnson V; six images with a total exposition time of 7500 sec for each target).
Fig.~\ref{fig:map} shows the location of the four target fields on the body of M31.
In  Table~\ref{tab:targets} we give the basic parameters of the target clusters. Their images in the F606W band are shown in Fig.~\ref{fig:targets}.

The pointings were arranged to contain as many clusters or candidate clusters
as possible because other confirmed or candidate GCs are present in the proximity
of the targets. This optimized the scientific output of the images.
For the fields with multiple objects, orientation requirements were set
to guarantee  the best coverage of all targets, and additional orientational constraints were added to the B292 pointing to avoid a bright star in the field of view.  
The field of B058  also contains B049 (already studied in Perina et al. 2009b); the field of B337 also contains B336  and in the field of B292 we discovered a new cluster that was never detected before (dubbed B531, according to the RBC nomenclature). For these clusters it was possible to obtain meaningful CMD that are presented and discussed below.
In the field of B058 we identified another three candidate GCs listed in the RBC (SK087B, SK088B, SK089B). These are faint and very compact objects for which we provide only a re-classification based on visual inspection of the ACS images.   
In  Table~\ref{tab:others} are listed some basic parameters of these additional objects, and their 
images in the F606W band are shown in Fig.~\ref{fig:others}. 

The data reduction was performed in the same way as in Perina et al.~(\cite{P09b}; P09b hereafter) on the pre-reduced images provided by STScI, 
using the ACS module of DOLPHOT
\footnote{See http://purcell.as.arizona.edu/dolphot/.} (Dolphin ~\cite{dol_a}),
a point-spread function fitting package specifically devoted to stellar photometry 
from HST images. The package identifies the sources above a fixed flux threshold on
a stacked image and performs the photometry on individual frames, accounts for
the hot-pixel and cosmic-ray masking information attached to the 
observational material, automatically applies
the correction for the charge transfer efficiency (CTE, Dolphin ~\cite{dol_b}) 
and transforms instrumental magnitude to the VEGAMAG and standard BVI system 
using the transformations by Sirianni et al.~(\cite{siria}). 


\begin{table*} 
 \centering
 \caption{Ages from integrated spectroscopy/photometry for the main targets.}
\label{tab:age}
\begin{tabular} {lccccc} 
  \hline
    &\multicolumn{5}{c}{}\\
ID  &\multicolumn{5}{c}{AGE [Gyr]$^a$}\\						                         
    &\multicolumn{5}{c}{}\\
 \hline
    &\multicolumn{1}{c}{P05} &\multicolumn{1}{c}{Be05+BC03}&\multicolumn{1}{c}{Be05+T03} &\multicolumn{1}{c}{W10} &\multicolumn{1}{c}{F10}\\
 \hline
    &\multicolumn{4}{c}{}\\
  B058-G119 & 6.4$\pm$4.1 & $\dots$& $\dots$ &2.02$\pm$0.10 & 4.000$\pm$1.325  \\  
  B292-G010 & 9.2$\pm$3.3 & 2.7$\pm$1.2 &  5.9$\pm$3.4    &1.00$\pm$0.10 & 0.003$\pm$0.000 \\  
  B337-G068 & 4.9$\pm$2.9 & 2.6$\pm$1.9 &  6.6$\pm$3.2        &2.03$\pm$0.10 & 1.900$\pm$0.360\\ 
  B350-G162 & 9.3$\pm$2.3 & 9.8$\pm$2.5 & 12.4$\pm$4.7    &1.99$\pm$0.10 & 1.278$\pm$0.148 \\  
     &\multicolumn{4}{c}{}\\
 \hline
\end{tabular} 
{\begin{flushleft}{($^{a}$) Ages from Puzia et al.~(\cite{puzia}; P05); Beasley et al.~(\cite{beas05}; Be05); Wang et al.~(\cite{wang10}; W10);
Fan et al.~(\cite{fan10}; F10). Beasley et al.~(\cite{beas05} derived two set of age estimates by comparing the same spectra with two different sets of SPSS models by Bruzual \& Charlot~(\cite{bc03}; BC03) and Thomas et al.~(\cite{thomas03}; T03); we report both estimates in this table.
Caldwell et al.~(\cite{C11}) do not provide an age estimate for these clusters, but assign a conventional value of 14 Gyr to all of them.
}
\end{flushleft}}
\end{table*}


\begin{table*} 
 \centering
 \caption{Spectroscopic metallicities and [$\alpha$/Fe] for the main targets .}
\label{tab:met}
\begin{tabular} {l|ccccc|c} 
  \hline
    &\multicolumn{5}{|c|}{}& \\
ID  &\multicolumn{5}{c|}{[Fe/H]$^a$}							                         &\multicolumn{1}{c}{[$\alpha$/Fe]} \\
    &\multicolumn{5}{|c|}{}& \\
 \hline
    &\multicolumn{1}{c}{P05$^{\ast}$} &\multicolumn{1}{c}{G09}  &\multicolumn{1}{c}{C11}&\multicolumn{1}{c}{B00} &\multicolumn{1}{c|}{F10} &\multicolumn{1}{c}{P05}		 \\
 \hline
    &\multicolumn{5}{|c|}{}& \\ 
  B058-G119$^c$ & -0.25$\pm$0.17 & -1.02$\pm$0.21 & -1.05$\pm$0.11 & -1.45$\pm$0.24$^b$ & -0.916$\pm$0.099 & -0.34$\pm$0.15   \\  
  B292-G010     & -1.63$\pm$0.49 & -1.54$\pm$0.37 & \dots          & -1.42$\pm$0.16     & -2.249$\pm$0.043 &  0.11$\pm$0.31   \\  
  B337-G068     & -0.59$\pm$0.11 & -1.08$\pm$0.15 & -1.17$\pm$0.14 & -1.09$\pm$0.32     & -0.433$\pm$0.113 &  0.00$\pm$0.11   \\ 
  B350-G162     & -1.81$\pm$0.37 & -1.54$\pm$0.31 & -1.43$\pm$0.13 & -1.47$\pm$0.17     & -1.029$\pm$0.127 &  0.38$\pm$0.28   \\  
    &\multicolumn{5}{|c|}{}& \\ 
 \hline
\end{tabular} 
{\begin{flushleft}
{($^{a}$) [Fe/H] from Puzia et al.~(\cite{puzia}; P05); Galleti et al.~(\cite{G09}; G09); Caldwell et al.~(\cite{C11}; C11); Barmby et al.~(\cite{barm00}; B00);
Fan et al.~(\cite{fan10}; F10).}\\
{($^{b}$) from Huchra et al.~(\cite{huch91}).}\\
{($^{c}$) Rich et al.~(\cite{rich05}) provides a CMD-based estimate of $[Fe/H]=-1.40\pm0.20$ for the cluster B058.}\\ 
{($^{\ast}$) Trasformed from [Z/H] with the formula of Thomas, Maraston \& Bender ~(\cite{thomas03}).}
\end{flushleft}}
\end{table*} 

%

We fixed the threshold for the search of sources on the images at 3$\sigma$ above
the background. DOLPHOT provides as output the magnitudes and positions of the
detected sources as well as a number of quality parameters for a suitable sample
selection, with regard to of the actual scientific objective one has in mind.

Owing to the adopted peculiar pointing strategy described above, we were forced to reduce
the F435W and F606W frames of each cluster independently. The output positions
were corrected for the ACS geometrical distortions and the F435W and F606W were then cross-correlated to match common sources\footnote{The match was performed with CataXcorr, a code aimed at cross-correlating catalogs and finding astrometric solutions, developed by P. Montegriffo at INAF - Osservatorio Astronomico di Bologna, and successfully used by our group for the past 10 years.}.
This non-standard procedure prevented the DOLPHOT routine from calculating the required color term
to perform the transformations from instrumental magnitudes to standard B,V magnitudes automatically; we made this final passage a-posteriori with the transformations by Sirianni et al.~(\cite{siria}). Below we will always consider only the photometry in the Johnson B,V passbands. Similarly to P09b, we retained in our final catalogs all  the sources with valid magnitude measurements in both passbands, global  quality flag = 1 (i.e., best measured stars), 
{\em crowding}  parameter $\le 0.5$, $\chi^2<2.5$, and {\em sharpness} parameter  between -0.3 and 0.3 (see Dolphin ~\cite{dol_b} for details on the meaning of the various quality parameters). This selection cleans the sample from the majority of spurious and/or bad  measured sources without significant loss of information and was found  to be appropriate for the
whole data set.

The  limiting  magnitudes  of our  photometry  range  from  V$\sim$27.0 for
the field of B058, which is projected 
onto a much more crowded background compared to the other clusters (see Figs.~\ref{fig:map} and \ref{fig:targets}), to  V$\sim$29.0 for the other fields. 
The internal photometric errors of individual measures are in
the range 0.005 - 0.08 mag for stars brighter than V=27 (see the
error bars in Figs.~\ref{fig:isoc1}, \ref{fig:isoc2}, and \ref{fig:cmd_others}).

\subsection{Artificial stars experiments}

We performed artificial stars experiments (generating $\sim 10^5$ fake stars per field) to study in detail  the
completeness of the samples as a function of magnitude, color, and distance from the 
cluster's center (i.e., as a function of crowding), using the automated procedure included in  DOLPHOT (see P09b for details). 

The fields of B292, B336, and B350 all have completeness $\ga$80\% for V$\la$27, more 
than sufficient to detect a blue HB where it is present. On the other hand, the TO stars 
are located in a CMD region in which the completeness is less then 50\%, which prevents us from reaching 
any suitable conclusion from this CMD feature.
The field of B058 is projected onto 
the so-called 10 kpc ring, a site of ongoing star formation in the thin disk of M31
where the stellar density (and then the crowding) is much higher, conseguently the completeness is significantly worse 
($\ga$80\% for V$\la$24.5) and barely reaches the level of the red HB.

\subsection{Classification of secondary targets}

The CMD of the secondary target B049 obtained from these data is very similar to that already published and discussed in P09b, which was obtained from a different dataset; 
we refer the interested reader to this previous publication for details.

SK087B is the faintest object in our sample and is located in the dense disk field of B058. 
The CMD of the stars located within a radius of $~\sim2.5\arcsec$ appears to be indistinguishable 
from that of the surrounding field and the same stars do not show any over-density with respect to the field. For these reasons re-classified this object as an asterism. The high-resolution ACS images clearly reveals that SK088B is not a genuine cluster but a bright star, instead.
SK089B is imaged only on the frame F606W, so we cannot provide its CMD, however, it appears to be 
an OB association in the proximity of its dust cocoon. We maintained the classification of this object.


\begin{table} 
\centering
\caption{Fundamental parameters and classification for additional clusters and candidate clusters  included in the images.}
\label{tab:others}

\begin{tabular} {l c c c c c} 
  \hline
\\
ID  & RA (J2000) & Dec (J2000) &  V$_{int}$ & E(B-V)$^\ast$ & type$^{\ast\ast}$ \\
    &            &             &    &               &             \\
 \hline                                                                                                                                                  
\\
  {\tiny B049}       & {\tiny 00 41 45.6} & {\tiny+40 49 53.7} &  17.56 & {\tiny 0.16$\pm$0.02} &1(1) \\	     
  {\tiny SK087B}          & {\tiny00 41 48.2} & {\tiny+40 47 31.5} &  18.83 & $\dots$       &7(2) \\		
  {\tiny SK088B}          & {\tiny00 41 51.5} & {\tiny+40 47 06.6} &  16.98 & $\dots$       &6(2) \\		   
  {\tiny SK089B}          & {\tiny00 41 54.7} & {\tiny+40 47 14.6} &  18.57 & $\dots$       &2(2) \\	     
  {\tiny B531}            & {\tiny00 36 14.7} & {\tiny+40 57 56.6} & 19.63$^{\star}$& $\dots$       &1    \\  
  {\tiny B336}       & {\tiny00 40 47.6} & {\tiny+42 08 41.4} &  17.81 & {\tiny 0.42$\pm$0.04} &1(1) \\		   
\\
 \hline
\end{tabular} 
{\begin{flushleft}
{V$_{int}$ from the RBC.}
{~~$^{\ast}$ from Fan et al.(~\cite{fan08,fan10}). Caldwell et al.~(\cite{C11}) reported E(B-V)= 0.32 and 0.05 for B049 and B336, respectively.}
{~~$^{\ast\ast}$: RBC Classification flag: 1- confirmed cluster; 2- gc candidate; 6- star; 7- asterism. The classification reported in the RBC before this study are enclosed in parentheses.}
{~~$^{\star}$ See Sect.~\ref{secB531}. 
}
\end{flushleft}}
{\begin{flushleft}
\end{flushleft}}
\end{table} 


   \begin{figure}
   \centering
   \includegraphics[width=9cm]{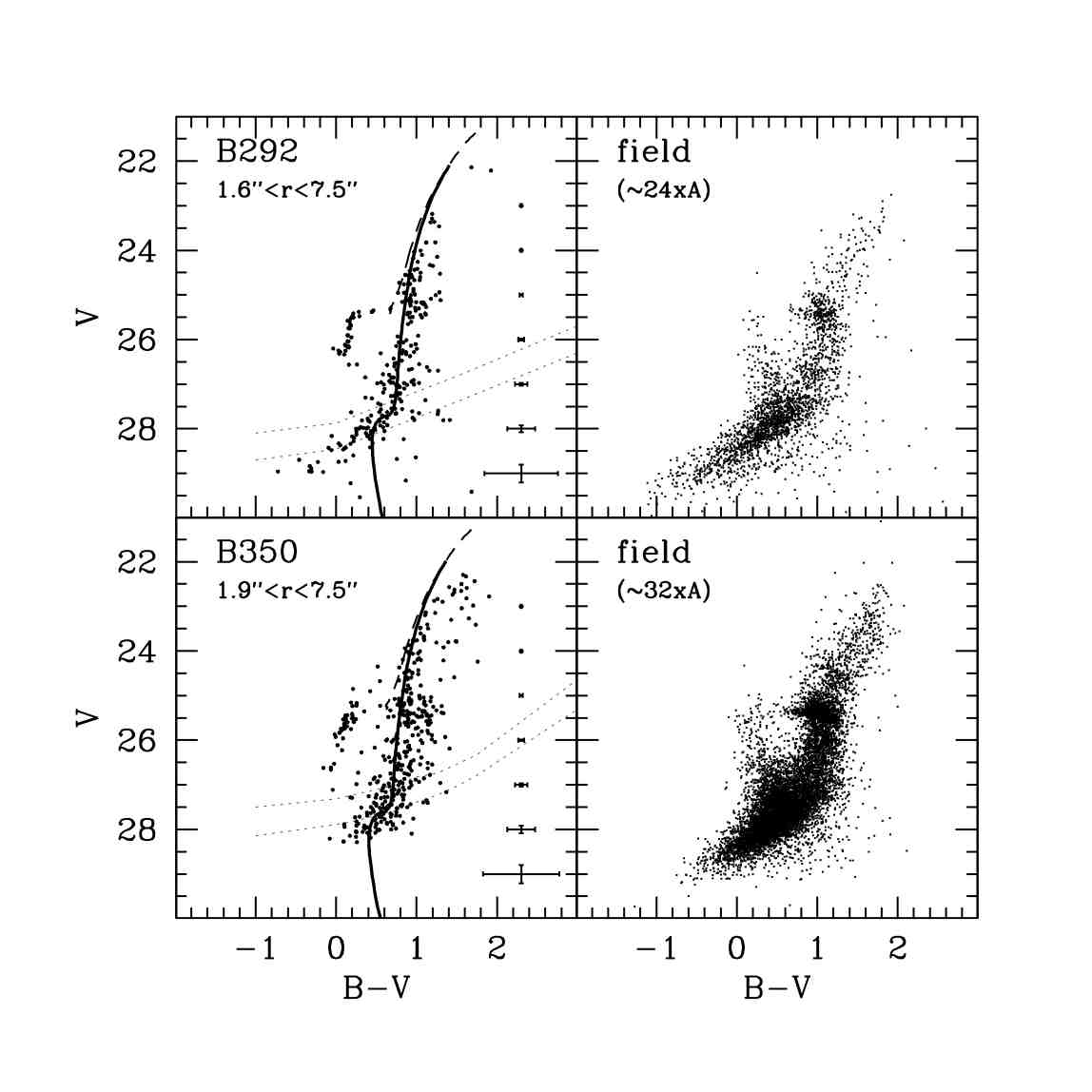}
      \caption{Left panels: CMD of the stars enclosed in the reported
      annulus for the target clusters B292 and B350. Overplotted on the CMD
      there are the 80\% and 50\% completeness levels (dotted lines) and an isochrone
      from the Girardi et al.(~\cite{isocPD}) dataset
      with age and metallicity corresponding to the spectroscopic estimates by 
      P05 (see Table~\ref{tab:age} and ~\ref{tab:met});
      the red giant branch is plotted with a thick continuous
      line and the asymptotic giant branch with a thin dashed line.  
      The reddening and distance moduli are taken from 
      Fig.~\ref{fig:templates}, below.
      Right panels: CMD of the field stars in the neighborhood of the cluster.
      The area of the considered field is reported in units of the area of the annulus
      shown in the left panels (A).}
         \label{fig:isoc1}
   \end{figure}
   \begin{figure}
   \centering
   \includegraphics[width=9cm]{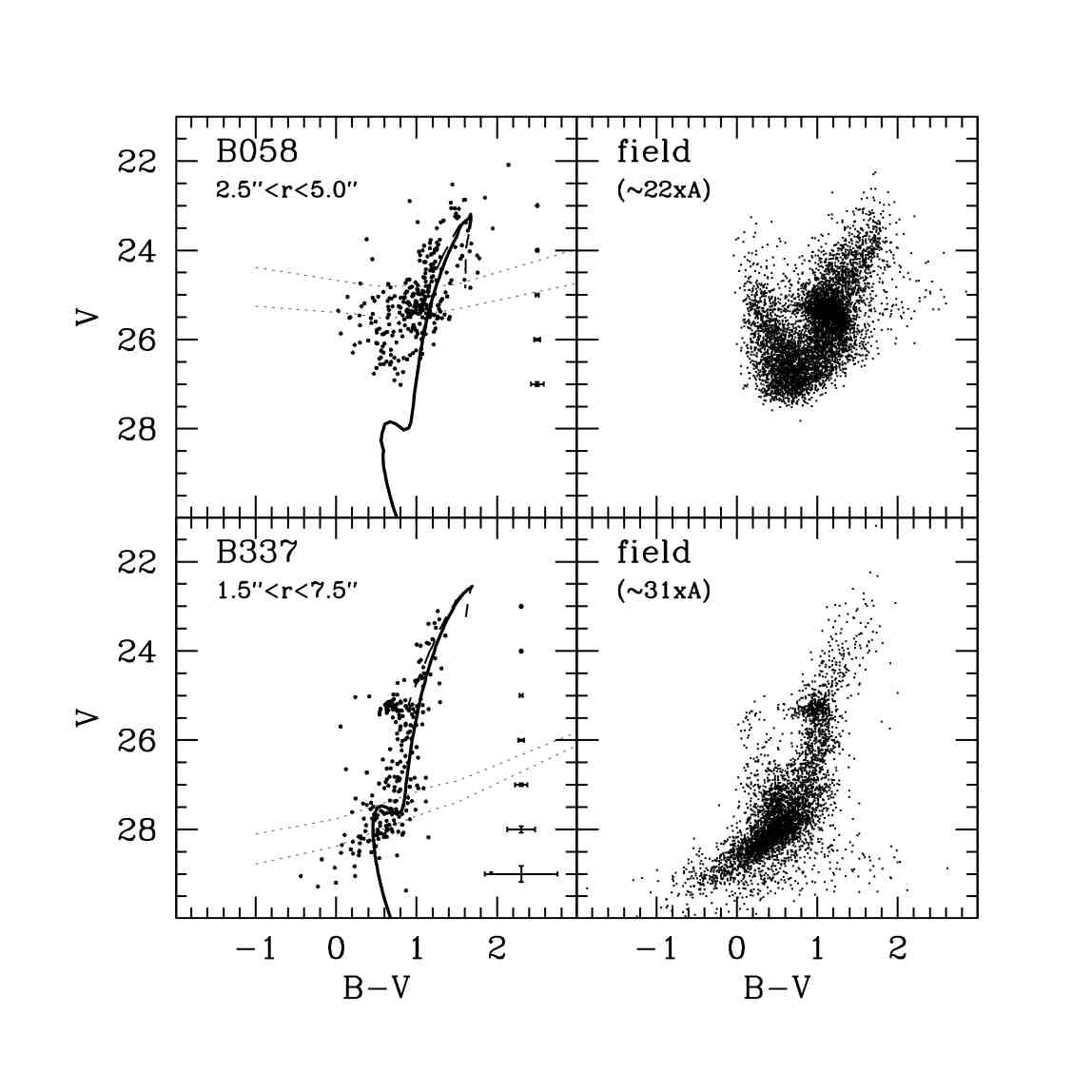}
      \caption{Same as \ref{fig:isoc1} for the other two target clusters
      B058 and B337.}
         \label{fig:isoc2}
   \end{figure}

   \begin{figure}
   \centering
   \includegraphics[width=9cm]{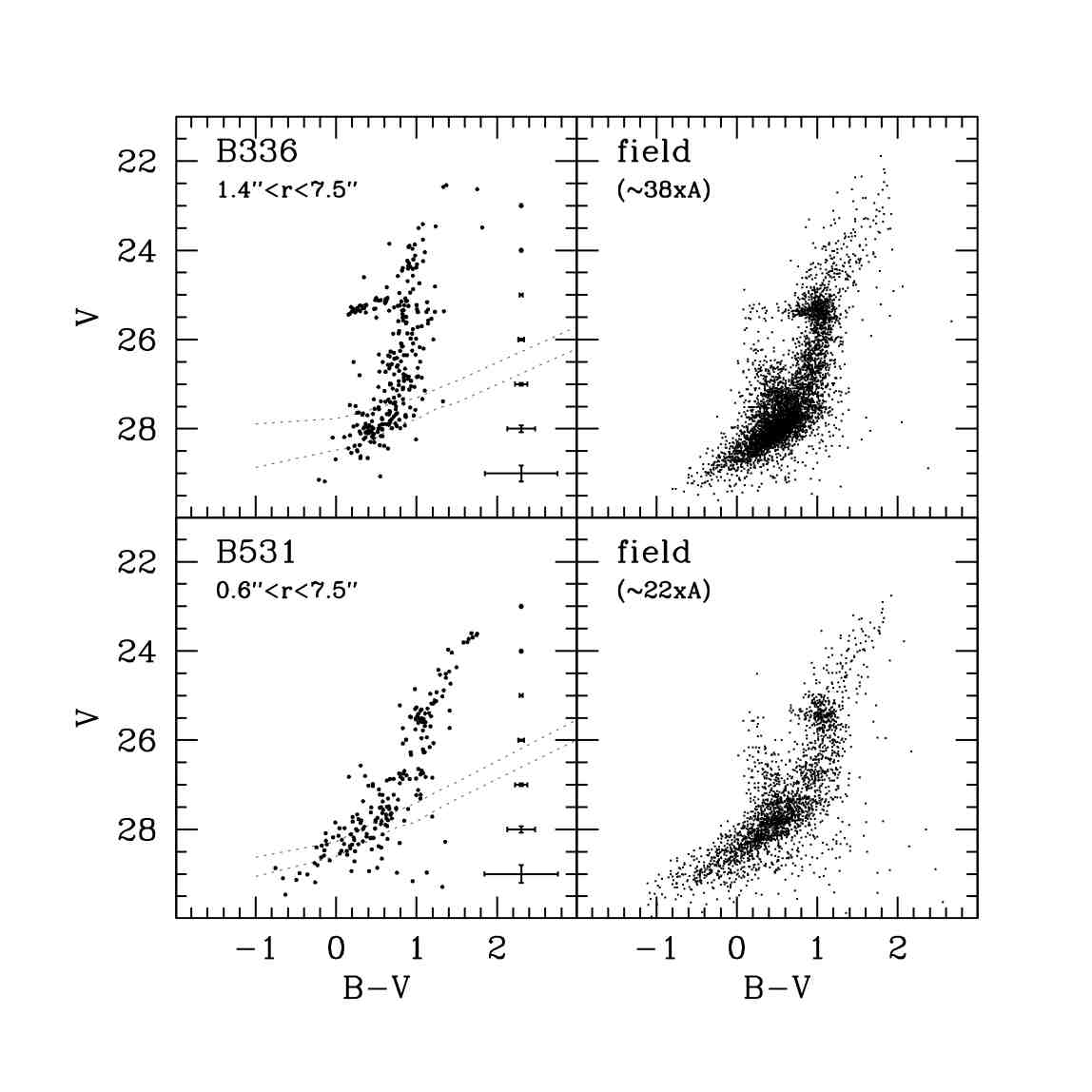}
      \caption{Same as Fig.~\ref{fig:isoc1} but for the clusters
      B336 and B531. The CMD of B049 is very similar to that obtained by P09b and 
      is not shown here.}
         \label{fig:cmd_others}
   \end{figure}

\section{The CMDs}
\label{cmd}

Below we present and discuss the CMDs of the four main target clusters, of the secondary target cluster B336 and of the newly detected cluster B531.
The individual CMDs are shown in Figs. \ref{fig:isoc1}, \ref{fig:isoc2} (target clusters), 
and \ref{fig:cmd_others} (secondary clusters). 
The cluster CMDs shown in the left panels sample the stellar population within an annulus around 
the cluster center. The inner limit of the annulus is set by the crowding level at the center, which prevents us from performing useful photometry in the most central region of the cluster, the outer limit is set by the need to avoid strong contamination by the 
surrounding field population. The inner and outer radii of the adopted annuli are indicated 
for each cluster. In the right panels we plot the CMD of the field adjacent to each cluster. The field population is sampled over a wide region outside the outer limit of the cluster and 
the area of the considered field is reported in units of the area of the annulus 
enclosing the cluster stars.

In the left panels of Figs.~\ref{fig:isoc1} and \ref{fig:isoc2} we over-plotted
the isochrones from the Girardi et al.(~\cite{isocPD}) set on the cluster CMDs, 
taking the age and the metallicity estimated by P05 for these clusters. The isochrones have been shifted to the same distance and reddening of the clusters adopting the values derived in Sect.~\ref{procedure} and listed in Table~\ref{tab:newparam}. 
Clearly, the CMDs are not deep enough to reliably test the region of the upper MS and MSTO that is expected for populations of that age. The handful of stars lying around the expected sub-giant branch (SGB) level are fully consistent with being largely caused by field contamination (see Fig.~\ref{fig:low_lim}). In any case, we can obtain strong and meaningful lower limits to the age of three of the four main targets (excluding B58) by searching for the oldest isochrone whose SGB clearly is {\em above} the observed candidate SGB stars (see Fig.~\ref{fig:low_lim}, below). The resulting limits, (age $> 6.3$~Gyr for B292, age $> 4.5$~Gyr for B337, and age $> 5.6$~Gyr for B350) are sufficient to rule out the estimates from integrated photometry by W10 and F10, which point to ages significantly younger than P05 (see Table~\ref{tab:age}). On the other hand,
we have to rely on evolved stars to constrain the age of our clusters in the range invoked by P05.

\begin{figure}
   \centering
   \includegraphics[width=11cm]{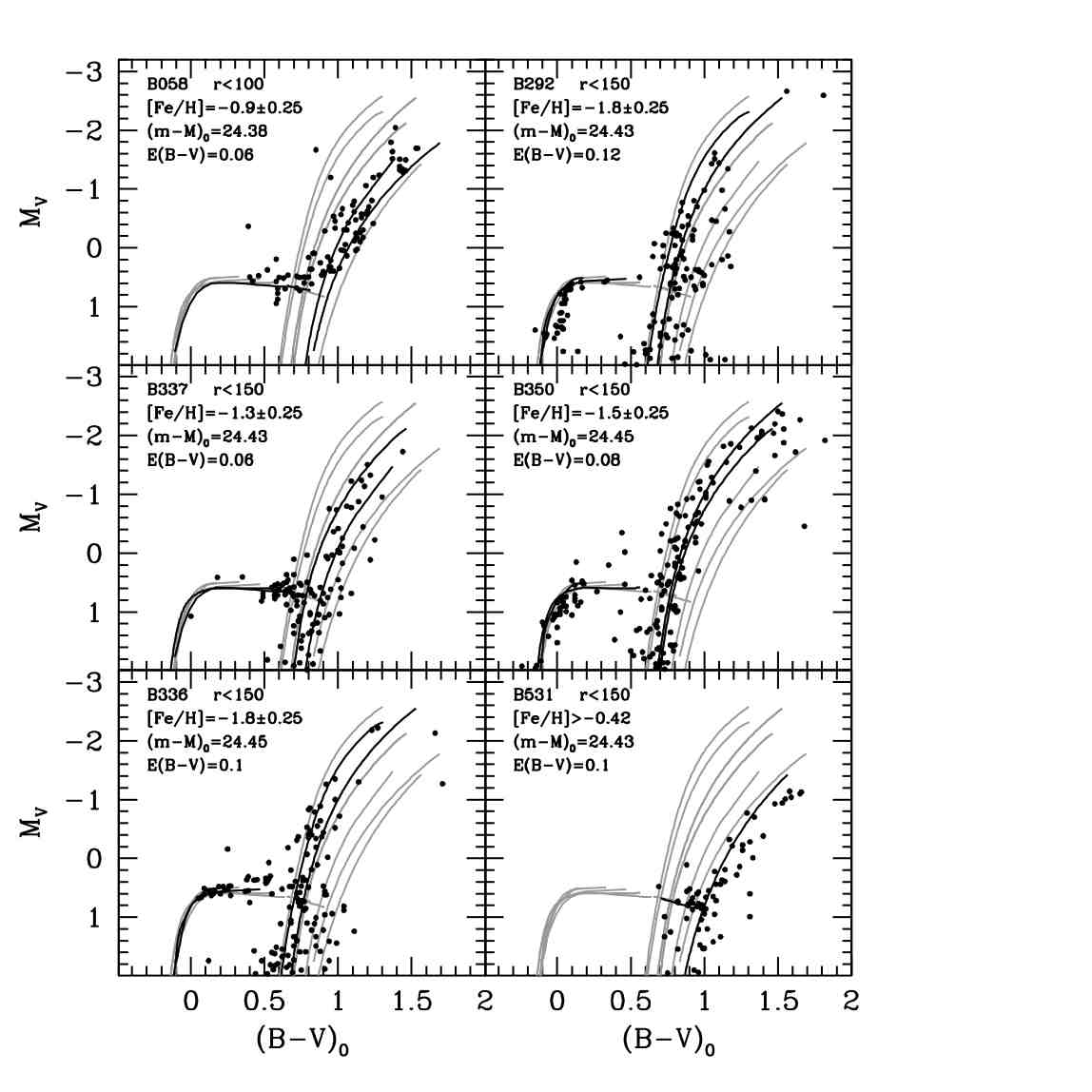}
      \caption{CMDs of the clusters after statistical decontamination from field
      stars with a superposed grid of HB and RGB templates. 
      The ridge lines that provide the best fit to the observed HB and bracket most of the RGB population are plotted as heavier lines to highlight them. The adopted values for the metallicity, reddening, and distance modulus are reported in the upper left panel of each plot.}
         \label{fig:templates}
   \end{figure}

In most CMDs the evolved population of the cluster can be clearly  distinguished from the field (see also the de-contaminated CMD presented in Sect.~\ref{procedure}). 
In particular, B292 and B350 display an obvious blue HB that cannot be accounted for by field contamination (see Fig.~\ref{fig:templates} and Table~\ref{tab:hb}). As said, this is a characteristic feature of very old metal-poor GCs; its mere presence strongly argues against ages younger than 10-11 Gyr. For the same clusters, the over-plotted isochrones predict the (possible) presence of asymptotic giant branch stars brighter than the RGB tip (dashed line) while no such stars are observed, providing further support to the idea that these clusters are indeed older than the considered models. It is worth to stress here that for B292 and B350 an age of 12.5 Gyr (i.e., typical of old MW GCs, Dotter et al.~\cite{dotter}) is within 1$\sigma$ from the age estimates by P05, hence fully compatible with them (see Table~\ref{tab:age}). It may be interesting to investigate why the accurate Lick indices analysis performed by those authors {\em singled out} these clusters as being younger than the bulk of the others. We will attempt to achieve some insight into this question in Sect.~\ref{discu}. On the other hand, even an age $\simeq 10$~Gyr would strongly disagree with the values provided by W10 and F10, suggesting that these estimates are not only systematically too low, but appear to have unrealistically low associated errors (typically lower than 1 Gyr; Table~\ref{tab:age}).


\begin{table}
\centering
\caption{Reference grid of template Galactic globular clusters.}
\label{tab:templates}
\begin{tabular}{@{}lcccc@{}}
\hline
\\
 ID & [Fe/H] & E$(B-V)$ & $(m-M)_{0}$ \\
    &        &  mag    &  mag     \\
\hline
\\
NGC7078 (M15)  & --2.16 & 0.10 & 15.37   \\
NGC5824        & --1.87 & 0.13 & 17.93   \\
NGC6205 (M13)  & --1.65 & 0.02 & 14.48   \\
NGC5904 (M5)   & --1.40 & 0.03 & 14.46   \\
NGC6723        & --1.12 & 0.05 & 14.85   \\
47 Tuc         & --0.71 & 0.04 & 13.37   \\
NGC6624        & --0.35 & 0.28 & 15.36   \\
\\
\hline
\end{tabular} 
{\begin{flushleft}{NOTES: Metallicities are from Zinn (1985); all other parameters are from 
Harris (1996) (online update  2003).
B,V photometry is from Piotto et al. (2002).}
\end{flushleft}}
\label{tab:ggc1}
\end{table} 

\subsection{Determination of the cluster parameters}
\label{procedure}

Before proceeding with the analysis of the cluster properties, 
we applied the field decontamination
procedure described in Bellazzini et al.~(\cite{bel99}). This method
is based on a clipping routine which, making use of the local
density on the CMDs of the field and of the cluster, computes
the probability that a given star is a member of the cluster and
retains or rejects stars from the cluster CMD according to that
(see P09b for further details).

Assuming that the clusters are classical old globulars,
we derived a simultaneous estimate of the distance modulus, reddening, and metallicity 
by comparing them with a set of CMD templates of
well studied Galactic GCs, exactly in the same way as in P09b. 
We searched for the set of parameters ($(m-M)_{0}$ ,
E(B-V) and [Fe/H]) producing the best match between the
observed RGBs and HBs and the ridge lines of the template
clusters in the absolute plane, given the direction of the
reddening vector $A_{B} = 4.145E(B-V)$, $A_{V} = 3.1E(B-V)$
(Schlegel et al.~\cite{sch98}).
The best match was judged by eye guided by (extensive)
experience. Color and magnitude shifts are applied iteratively 
until a satisfactory match with any RGB and
HB template is found. From these shifts we obtained estimates of
the reddening and distance, while the metallicity was estimated by
interpolation between the two RGB ridge lines bracketing the
observed RGB locus.
As starting values for the iterative procedure we used
E(B-V) = 0.08 for the foreground reddening (Barmby
et al.~\cite{barm07}; Burstein \& Heiles~\cite{BH84}) and the distance modulus
$(m-M)_{0} = 24.47$ (McConnachie et al.
~\cite{mcc05}). The ridge lines of the reference GGCs were derived
from publicly available photometry (Piotto et al.~\cite{pio02}) and were shifted to
the absolute reference frame by correcting for reddening and
distance using the values listed in Table~\ref{tab:ggc1}. These reference GGCs
were chosen to provide a sufficiently fine and regular 
sampling over a wide enough range of metallicities for a correct
characterization of the target GCs.

In Fig.~\ref{fig:templates} we show the field-decontaminated CMDs
with the reference grid of GGC ridge lines over-plotted. 
The values of metallicity, reddening, and distance corresponding to the
best match are also reported in each individual panel, as well
as in Table~\ref{tab:newparam}; the typical uncertainty on the distance modulus is 
0.2 mag, 0.04 mag in E(B-V), and 0.25 dex in metallicity.
In Sect.~\ref{individual} we briefly discuss each individual cluster. 


\begin{table} 
 \centering
 \caption{Cluster parameters from the analysis of the CMD.}
\label{tab:newparam}
\begin{tabular} {l c c c c c} 
  \hline
\\
ID & E(B-V) & $(m-M)_{0}$ & {[Fe/H]} \\
   & mag        & mag           &    \\
 \hline
\\     
  B292-G010 & 0.12 & 24.43 & -1.80$\pm$0.25    \\  
  B337-G068 & 0.06 & 24.43 & -1.30$\pm$0.25    \\ 
  B350-G162 & 0.08 & 24.45 & -1.50$\pm$0.25    \\
  B336-G067 & 0.10 & 24.45 & -1.8 $\pm$0.25    \\
  B531      & 0.10 & 24.43 &      $>$-0.42     \\  
 \\
 \hline
\end{tabular} 
\end{table} 


\begin{table} 
 \centering
 \caption{Star counts in CMD boxes enclosing the BHB for the three clusters B292, B350, and B336. 
 }
\label{tab:hb}
\begin{tabular} {l c c c c c} 
  \hline
\\
ID & $N_{obs}^{GC}$ & $N_{obs}^{field}$ & $N_{exp}^{GC}$ & $\frac{N_{obs}^{GC}-N_{exp}^{GC}}{\sigma}$ & $\frac{Area_{GC}}{Area_{field}}$ \\
\\
 \hline
\\      
  B292-G010 & 30 & 25 & 1.05 $\pm$ 0.21 & 5.3 & 0.042 \\  
  B350-G162 & 37 & 76 & 2.36 $\pm$ 0.27 & 5.7 &	0.031 \\
  B336-G067 & 23 & 21 & 0.55 $\pm$ 0.12 & 4.7 &	0.026 \\
 \\
 \hline
\end{tabular} 
{\begin{flushleft}
$N_{exp}^{GC}$ is the number of cluster stars expected
 in a given box from the field population, computed by rescaling the observed counts in the
 sampled field by the ratio between the sampled GC and field areas (last column).
 In the fifth column we list the background-subtracted star counts in the GCs sampled areas
 in unit of $\sigma$.
\end{flushleft}}
\end{table} 

\subsection{Results for individual clusters}
\label{individual}

\subsubsection{B292=G010}
The cluster B292 lies at a fairly large projected distance from the center of M31, in a low-density region
where the contamination by the field stars is very low. Its CMD (Fig.~\ref{fig:isoc1}, top-left panel), is characterized
by a steep RGB and a well defined and populated BHB that is obviously associated with the cluster:
star counts within a box enclosing the feature indicate an overabundance of BHB stars
in the cluster with respect to the surrounding field (see Table~\ref{tab:hb}) at 5 $\sigma$ level.

The best match with the Galactic RGB and HB ridge lines is obtained by assuming a value of reddening 
$E(B-V)=0.12$ mag and a distance modulus of $(m-M)_{0}=24.43$.
With these assumptions, the cluster's RGB fall 
between the ridge lines of NGC5824 and M13, indicating that 
the metallicity of B292 is $[Fe/H]=-1.8\pm0.25$.
This value agrees well with the spectroscopic metallicity estimates by P05 and Galleti et al.~(\cite{G09}, hereafter G09), while it is just marginally more metal-poor than what was found by Barmby et al.~(\cite{barm00}, hereafter B00), and 0.45 dex more metal-rich than the value obtained by
F10 from integrated photometry. 
The mismatch in color between the vertical parts of observed and template BHB may suggest a solution with a higher reddening and, consequently, a lower metallicity; 
this would not change the main conclusions of the present analysis significantly. 
We preferred the solution presented in Fig.~\ref{fig:templates}, which gives greater weight to 
the five stars at intermediate color ($0.1\le (B-V)_0\le 0.6$, around $M_V\simeq 0.6$), because, 
if interpreted as genuine cluster members, they provide a constraint on the vertical direction 
(distance) that is otherwise lacking. The best match of the color of the vertical portion of 
the HB can be obtained by adopting E(B-V)=0.19, and $(m-M)_0=24.3$. This solution appears 
unlikely because it would imply that this cluster is located several kpc above (and in front of) 
the M31 disk while suffering from a degree of extinction quite typical of clusters embedded in 
the disk (see Perina et al.~\cite{ymc}). A small-scale Galactic cloud on this line of sight seems 
the only possibility to make this solution viable. In any case, this solution also yields a low 
metallicity ($[Fe/H]\simeq -1.9$) and the old age implied by the presence of BHB.

   \begin{figure}
   \centering
   \includegraphics[width=11cm]{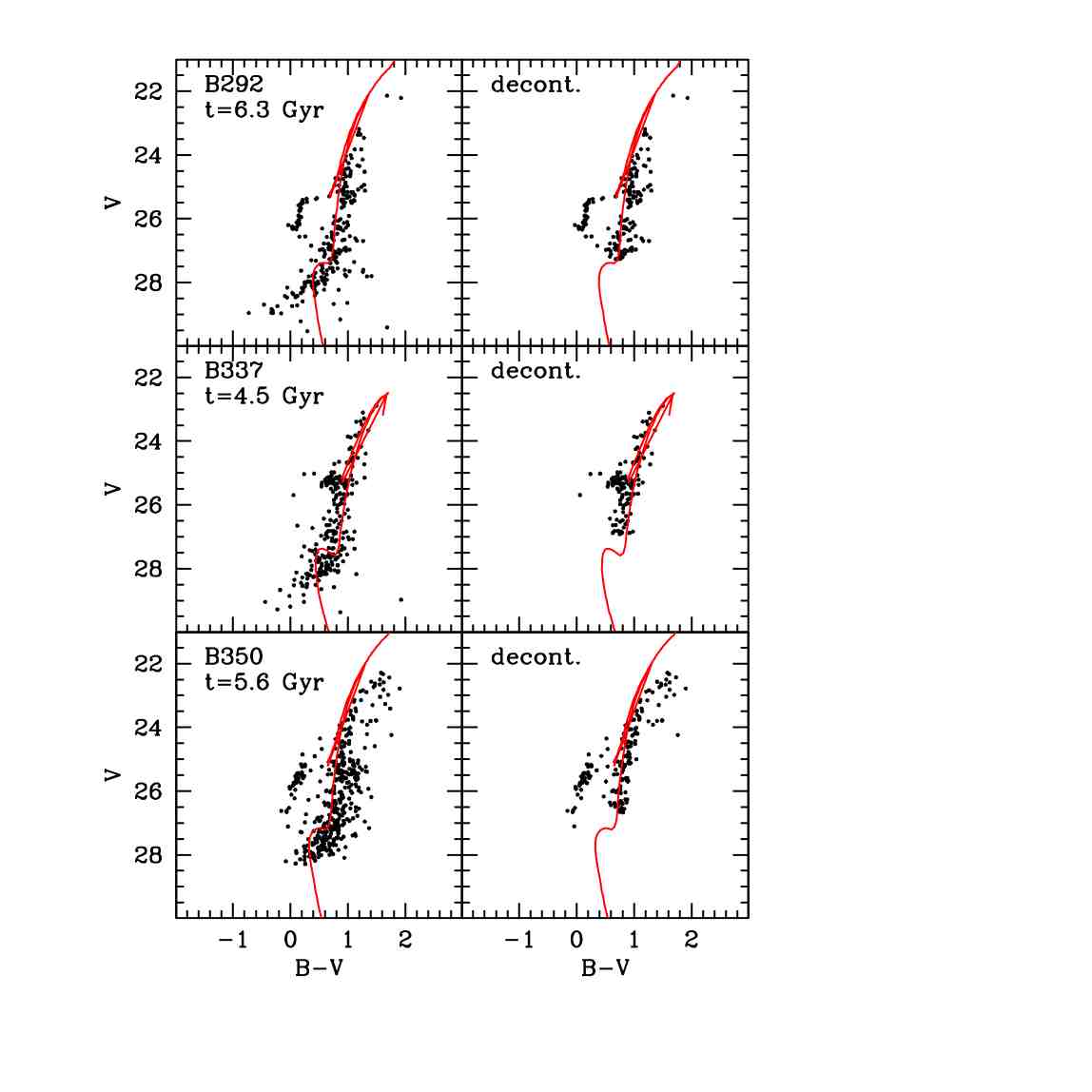}
      \caption{CMDs of the candidate intermediate-age clusters (stars selected in the same annuli as in Figs.~\ref{fig:isoc1} and \ref{fig:isoc2}) before (left panels) and after (right panels) the statistical decontamination from field stars. We superposed on the CMDs the oldest-age isochrones whose upper MS and SGB trace the blue/bright edge of the observed stars to set a strong lower limit to the age of the clusters (metallicity, reddening, and distances are the same as those adopted in Figs.~\ref{fig:isoc1} and \ref{fig:isoc2}). The age of the adopted isochrone is reported as a label in the upper-right corner of the left panels.
      Note, however, that there are no such faint stars in the de-contaminated CMD, suggesting that they are not associated to the clusters.}
         \label{fig:low_lim}
   \end{figure}

\subsubsection{B350=G162}

The CMD of B350 (Fig.~\ref{fig:isoc1}, bottom-left panel) shows a well populated blue HB 
clearly associated with the cluster (see Table~\ref{tab:hb} and Fig.~\ref{fig:low_lim}).
The RGB is steep, which indicates a low metal content.

The best match of the RGB and HB features with the corresponding Galactic ridge lines is obtained by assuming
a distance modulus of $(m-M)_{0}$ = 24.45 and maintaining the adopted starting value 
of $E(B-V)=0.08$ mag for the reddening, in good agreement with what was reported by F10 (Table~\ref{tab:targets}).
With these assumptions, the cluster RGB fall between the ridge lines of M13 and M5, 
indicating for B350 a metallicity of $[Fe/H]=-1.5\pm0.25$. This is compatible with most of the values available in literature, except for the marginal discrepancy with the estimate by F10 from integrated photometry (see Table~\ref{tab:met}).

\subsubsection{B058=G119} 

Owing to the high degree of crowding, the CMD of B058 is not of sufficient quality to obtain a 
truly reliable solution. The CMD previously obtained by Rich et al.(~\cite{rich05}) from WFPC2 
data in the filters F555W (V) and F814W (I) with longer exposure times (5300~s and 5400~s, 
respectively) seems easier to interpret, because it appears to be slightly deeper and the RGB 
shape is straighter in the I-band. The solution presented here must be considered as only tentative, we suggest to keep the parameters by Rich et al.~(\cite{rich05}) as the best set. 

The 6.4 Gyr isochrone shown in Fig.~\ref{fig:isoc2} seems fully compatible with the observed CMD, 
but an old isochrone would fit equally well. The estimate by P05 had a large 
uncertainty ($6.4\pm4.1$ Gyr) and the CMD test is completely inconclusive. 
Ages as low as 4 Gyr or 2 Gyr, as proposed by F10 and W10, respectively, are compatible with 
the observed CMD as well.

\subsubsection{B337=G068}
B337 has a very clean CMD characterized by a steep RGB and a well populated red HB (see Fig.~\ref{fig:isoc2}).
The decontamination has little effect on the CMD of this cluster, which is positioned in a 
quite low-density region
(see Figs.~\ref{fig:templates} and \ref{fig:low_lim}).
  
The best match of these features with the corresponding ridge lines 
is obtained by assuming a value of reddening  of $E(B-V)=0.06$ mag (in excellent agreement with F10) and a distance modulus of $(m-M)_{0}=24.43$.
With these assumptions, the cluster RGB lies
between the ridge lines of M5 and NGC6723, indicating a metallicity of B337 of $[Fe/H]=-1.3\pm0.25$.
This value agree with the estimates by G09 and B00, while it is 
significantly lower than what was obtained by P05 and F10. (see Table~\ref{tab:met}). 
The CMD appears to be incompatible with this high metallicity, if genuinely old.

It is impossible in this case as well to draw a firm conclusion on the age of the cluster.
The red HB is bluer than that of the field, compatible with old metal rich 
population {\em \`a la} 47~Tuc. Three stars lying very close to the BHB templates are preserved 
in the de-contaminated CMD, but they are too few to be unequivocally attributed to the cluster. 
As said above, the faint end of the CMD seems already incompatible with the formal best-fit age 
proposed by P05 (4.9 Gyr; see Fig.~\ref{fig:low_lim}), but 
the tension relaxes if the uncertainties are taken into account (the same is true for the 
estimate of Beasley et al.~\cite{beas05}). On the other hand, the values proposed 
by F10 ($1.9\pm 0.4$ Gyr) and W10 ($2.0 \pm 0.1$ Gyr) are clearly too low to be compatible with the CMD.

\subsubsection{B336=G067}
B336 is a secondary target falling in the field of B337.
The CMD of B336 presented here shows a populous 
intermediate-blue HB population, implying an old age for the cluster (see Table~\ref{tab:hb}). 
This is also the only cluster in our sample that displays a population of candidate RR~Lyrae 
variables in significant excess with respect to the surrounding field, another indication of 
old age (Sect.~\ref{RR-Lyrae}). 
The best fit to the decontaminated CMD with the Galactic templates is obtained assuming a 
distance modulus of 24.45 and a
$E(B-V)=0.10$. The RGB of B336 lies between the ridge lines of NGC5824 and M13, indicating a metallicity  
$[Fe/H]=-1.8\pm0.25$ (see Fig.~\ref{fig:templates}).
Fan et al.~(\cite{fan10}) estimates from integrated photometry an age of $1.015\pm0.167$ Gyr, a value
incompatible with the observed CMD. The authors obtained a metallicity of $[Fe/H]-2.221\pm0.085$ in marginal disagreement with the 
value estimated here, and a reddening $E(B-V)=0.420\pm0.040$. The adoption of this high reddening 
value would shift all RGB and most BHB cluster stars to the blue of the bluest template, leading to an obvious inconsistency. Caldwell et al.~(\cite{C11}) report $[Fe/H]-2.49\pm0.59$ and $E(B-V)=0.05$, confirming that the cluster is not very reddened and is remarkably metal-poor.

\subsubsection{B531}
\label{secB531}
B531 is a newly detected cluster, found in the field of the target B292.
Its CMD in Fig.~\ref{fig:others} shows a red HB and an RGB quite tilted and redder then the reddest available Galactic template (NGC6624).
The fitting procedure of the decontaminated CMD with the Galactic templates 
provided the best result adopting a distance modulus of 24.43 and $E(B-V)=0.10$ mag (similar to 
what was found for B292 in the same field), indicating
B531 as a very metal-rich cluster with $[Fe/H]\ge-0.42$ (see Fig.~\ref{fig:templates}).
A solution with a higher reddening would lead to a lower metallicity estimate, but the match of 
the red HB level would be lost. Moreover, this line of sight is far from the visible disk of M31 
and is unlikely to be associated with high values of intrinsic extinction.
B531 lies at a projected distance $R_p\simeq 17$~kpc from the center of M31; the projection along 
the galaxy minor axis is $Y\sim 11$~kpc. In P09b it has been noted that metal-rich clusters 
located so far away from the galaxy center (and ``above'' the galaxy disk) are quite rare and 
may trace substructures associated to past accretion events in the M31 halo (see the case of 
B407, discussed in P09b, and also Mackey et al.~\cite{mack10}). 
If the high-metallicity of B531 will be confirmed, this cluster would be an interesting candidate ``accreted'' GC. A spectroscopic study, aimed at estimating the metallicity and the velocity of this cluster would be very valuable to follow-up this hypothesis. The integrated color of B531 $(B-V)_0=0.81$ is very similar to that of 47~Tuc [$(B-V)_0=0.84$, from Harris~\cite{harris}], thus supporting a (relatively) high metal content.

Integrated B,V, I  magnitudes were computed for this cluster by transforming aperture 
magnitudes obtained from Sloan Digital Sky Survey (Abazajan et al.~\cite{sdss}) calibrated 
images, as described in detail in Perina et al.~\cite{P09a} for VdB0. Adopting an aperture 
of radius 11~px$\simeq4.3\arcsec$, 
we obtained $g=20.08\pm0.04$, $r=19.32\pm0.04$, and $i=18.99\pm 0.04$. Lupton (2005) 
transformations\footnote{\tt www.sdss.org/dr4/algorithms/sdssUBVRITransform.html} were used to 
convert from g,r,i to B,V, I magnitudes.

   \begin{figure}
   \centering
   \includegraphics[width=9cm]{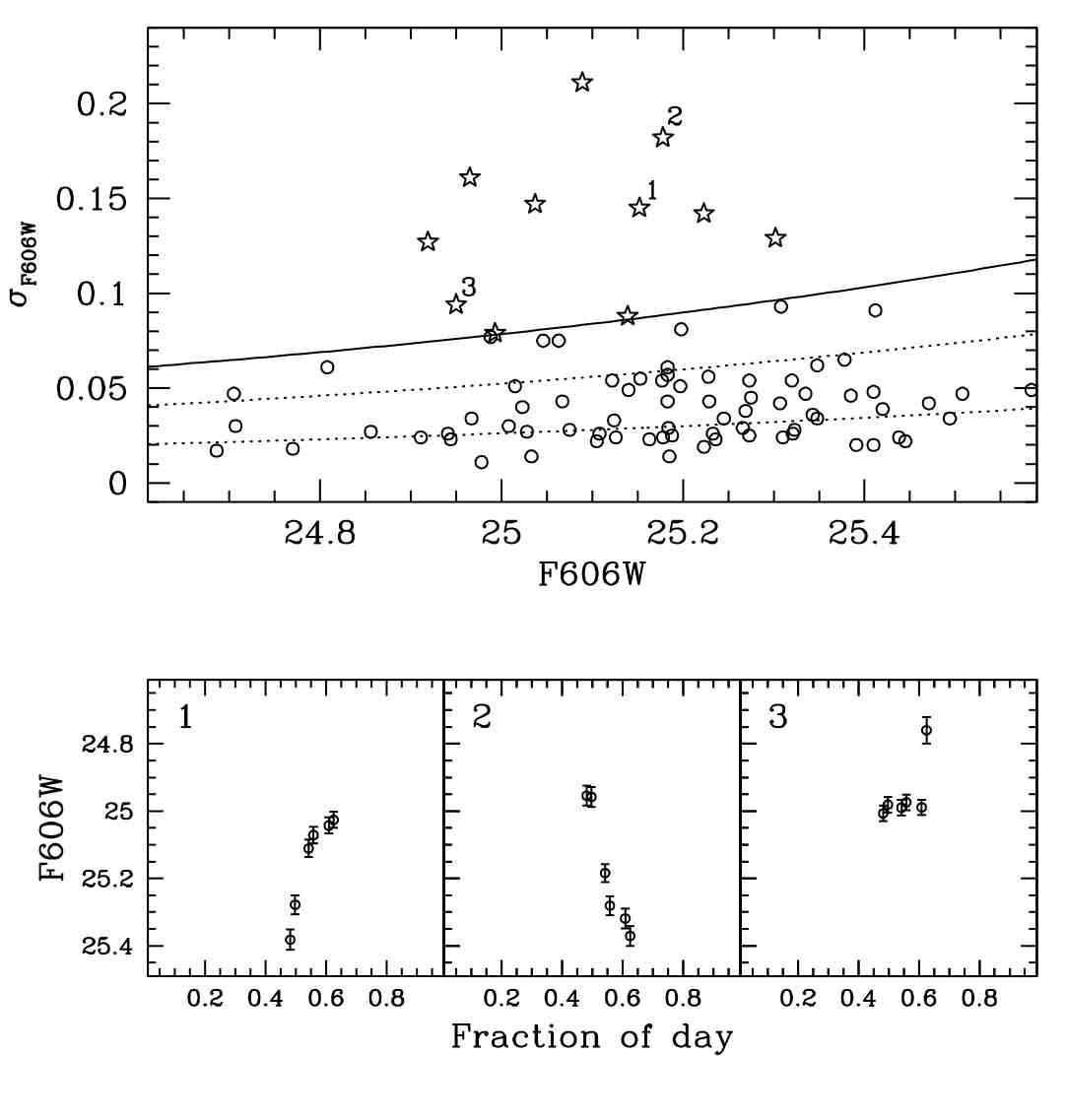}
      \caption{Process of detection and selection of candidate RR~Lyrae is illustrated using
      B336 as an example.
      Upper panel: the standard deviation of the five individual measures of F606W is plotted 
      as a function of the average F606W. The candidates RR~Lyrae were selected among stars with
      $24.9\le F606W\le 25.4$, as 
      stars with rms deviations larger then 3 $\sigma$ (open stars).
      The continuous line indicates the 3 $\sigma$ level and the dotted lines the 2 and 1 $\sigma$ levels.  
      Lower panels: Light curves of the candidates RR~Lyrae detected in the upper panel and labeled with 1, 2, and 3, taken as examples. 
      Candidates 1 and 2 show coherent patterns of variations at all sampled epochs, compatible with
      genuine RR~Lyrae variables. Candidate 3 is a classical case of spurious detection: F606W is constant at all epochs except one, likely corresponding to the hit of a cosmic ray over the PSF area of the star in the individual image at that epoch. }
         \label{fig:rrly1}
   \end{figure}

\section{Detection of RR~Lyrae variables}
\label{RR-Lyrae}

Another possibility to classify a cluster as very old is to identify a population of RR~Lyrae belonging to it. The considered observations were not planned for a search of variables and are far from optimal for this purpose. 
For each observed field, the observations in the two filters were acquired at a distance of two months, therefore they are not expected to sample the same phase range of any light curve. The four consecutive F435W images (per field) span an interval of $\sim 3.2$ hours (i.e., $\sim \frac{1}{4}$ of the typical period of a RR~Lyrae); the six consecutive F606W images (per field) span an interval of $\sim 4.0$ hours (i.e., $\sim \frac{1}{3}$ of the typical period of a RR~Lyrae).
Therefore, even if the amplitude of the variation is expected to be larger in F435W, the F606W observations are likely more efficient in detecting RR~Lyrae because of the larger number of exposures and larger phase coverage. 
 
With this observational material we can hope to detect ab-type RR~Lyrae variables with periods longer then $\sim5$~h only if they are caught during the rapid rise to maximum light or in the descending branch of their light curve, while they would go unnoticed if they are observed near their maximum/minimum. Moreover, because the exposures in the two filters are not consecutive, the detection of variability in both passbands is only a serendipitous event.
Similar limitations have been encountered also by Clementini et al.~(\cite{clem01}), who for the first time, 
searched for RR~Lyrae variables in a sample of M31 GCs using archive HST Wide Field Planetary Camera 2 (WFPC2) 
data. That study, however, proved the possibility of a successful
detection of likely candidates RR~Lyrae also with data not optimized for this type of investigation. It is important to stress that the present analysis is not intended to obtain a complete census of the RR~Lyrae population in the considered clusters but, instead, is aimed at verifying whether a significant population of RR~Lyrae can be identified in some of them.

\begin{figure}
   \centering
   \includegraphics[width=11cm]{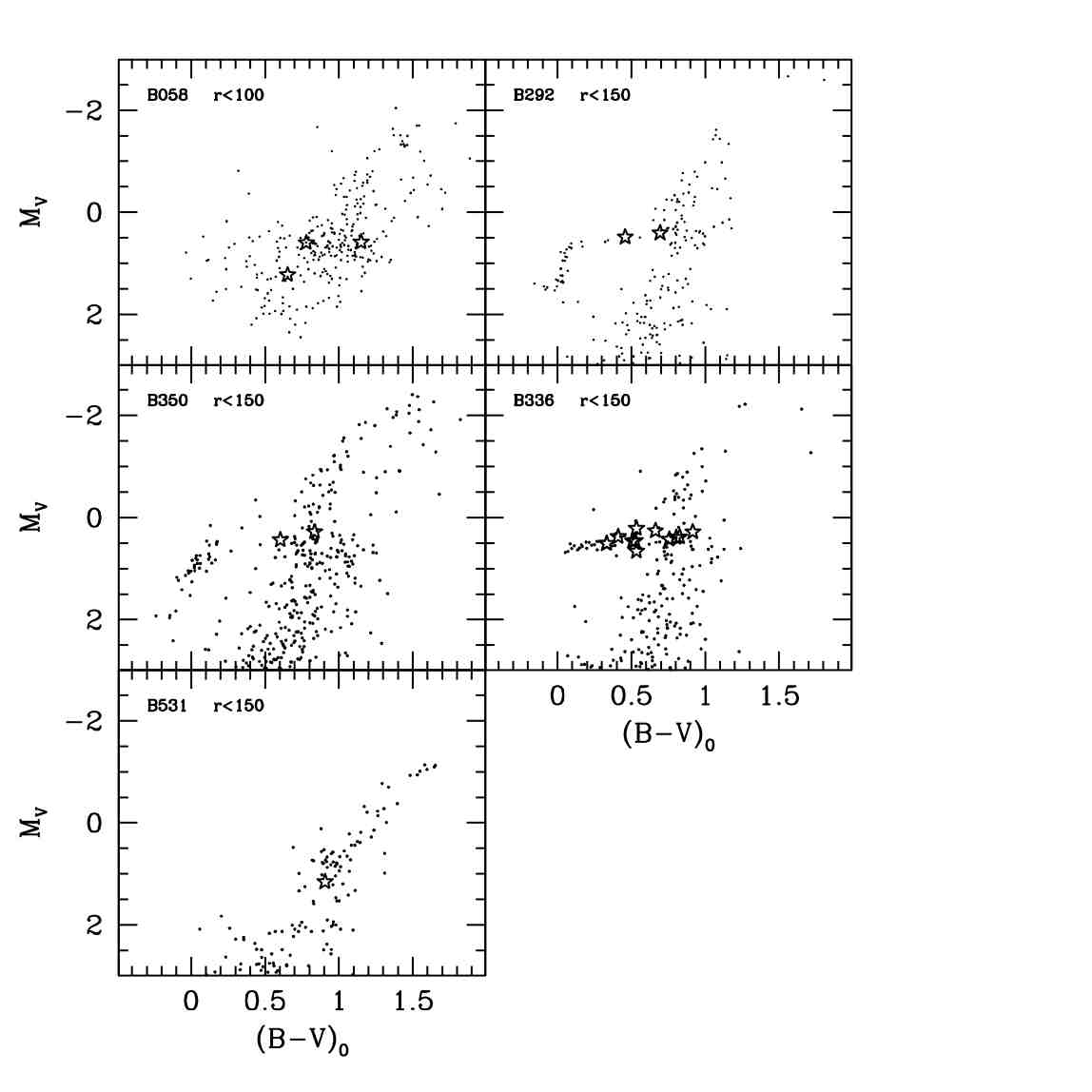}
      \caption{Candidate RR~Lyrae are overplotted as open stars on the cluster CMD.}
         \label{fig:rrly2}
   \end{figure}

To identify RR~Lyrae candidates, we adopted the following criteria, which are ilustrated in Fig.~\ref{fig:rrly1} :

\begin{enumerate} 
\item We searched for variability only in stars lying within $\pm1 mag$ of the HB level as estimated in Fig.~\ref{fig:templates}.

\item We considered objects with rms deviations in their estimate of the magnitude 
larger than or equal to three times the average deviation at their luminosity level.
Fig.~\ref{fig:rrly1} shows an example of application to the cluster B336: the selection procedure identifies
11 candidate variables (open star symbols) that satisfy this criterion.

\item For the selected stars the sequences of individual measurements were then displayed as a function of the epoch, searching for trends in the light curves that were compatible with RR~Lyrae type variations. 
Fig.~\ref{fig:rrly2} shows as an example the light curves of the three candidate variables labeled in Fig.~\ref{fig:rrly1}.
The stars labeled 1 and 2 show a coherent pattern of variations in all sampled epochs, 
compatible with genuine RR~Lyrae variables. On the other hand, candidate 3 is a classical case of spurious detection:
F606W is constant at all epochs except one, likely corresponding to the hit of a cosmic ray over the PSF area of the star in the individual image at that epoch.

\item We identified a star as a candidate RR~Lyrae variable if it satisfied all three criteria, at least in one of the two passbands.
\end{enumerate}

As a result of this analysis, we found 11 (B336), 3 (B058), 2 (B292, B350), 1 (B531) and 0 (B337) candidate RR~Lyrae stars in our six clusters, respectively. Their positions in the CMDs are shown in Fig.~\ref{fig:rrly2}.
We searched for candidate RR~Lyrae variables also in comparison fields surrounding the clusters
to quantify the probability that the variables detected in the cluster can be explained as 
merely due to the field population. Given the observed frequency of candidate variables in the 
surrounding fields, the number of field RR~Lyrae expected to fall on the cluster area are 0, for the 
clusters B336, B292, B531, B337, 1, for B350,  and 2 for B058. Hence, the only cluster for which a 
significant excess of RR~Lyrae is found with respect to the expectations is B336, providing further 
evidence supporting a very old age for this system.
The very red colors of most of the candidates in B058, B350, and B531 also cast some doubt on 
their classification as RR~Lyrae. Note that the lack of detection of a significant excess of 
candidate RR~Lyrae in the other clusters cannot be taken as evidence that no such variable is 
present there: our data are clearly not sufficient to reach this conclusion.

\section{Summary and discussion}
\label{discu}

The main aim of this study was to verify whether the intermediate age estimate obtained from integrated spectroscopy and/or photometry by several authors (Table~\ref{tab:age}) for the M31 clusters B058, B292, B337, and B350 was supported by the CMD obtained from the new HST-ACS observations. The final answer is completely inconclusive for B058 and uncertain for B337, even if in the last case there is marginal evidence for an older age compared to the dating from integrated light. On the other hand, both B292 and B350 show a significant and unequivocal population of blue horizontal branch stars, indicating that they are as old as the classical Galactic GCs (ages older than 11~Gyr; see Dotter et al.~\cite{dotter}, for a recent analysis and discussion). It should be noted that the age estimates provided by P05 are consistent within 1~$\sigma$ uncertainties with this conclusion. However, these clusters were singled out as likely younger than the bulk of M31 clusters at similar metallicity and this classification is clearly not confirmed here. On the other hand, ages as young as claimed by W10 and F10 (all $\le 4$~Gyr) are clearly ruled out by the observed CMDs for all main targets of the present study except (possibly) B058. 
The comparison with the results by Be05 significantly depends on the considered set of age estimates provided by these authors (see Table~\ref{tab:age}). It may be interesting to note that except for B350, their estimates based on the BC03 SPSS models are not compatible (too young) with the observed CMDs, even taking into account the reported uncertainties; on the other hand, the estimates obtained from the T03 models are similar to those by P05, and, in particular, the old age reported for B350 fully agrees with our results.

   \begin{figure}
   \centering
   \includegraphics[width=9cm]{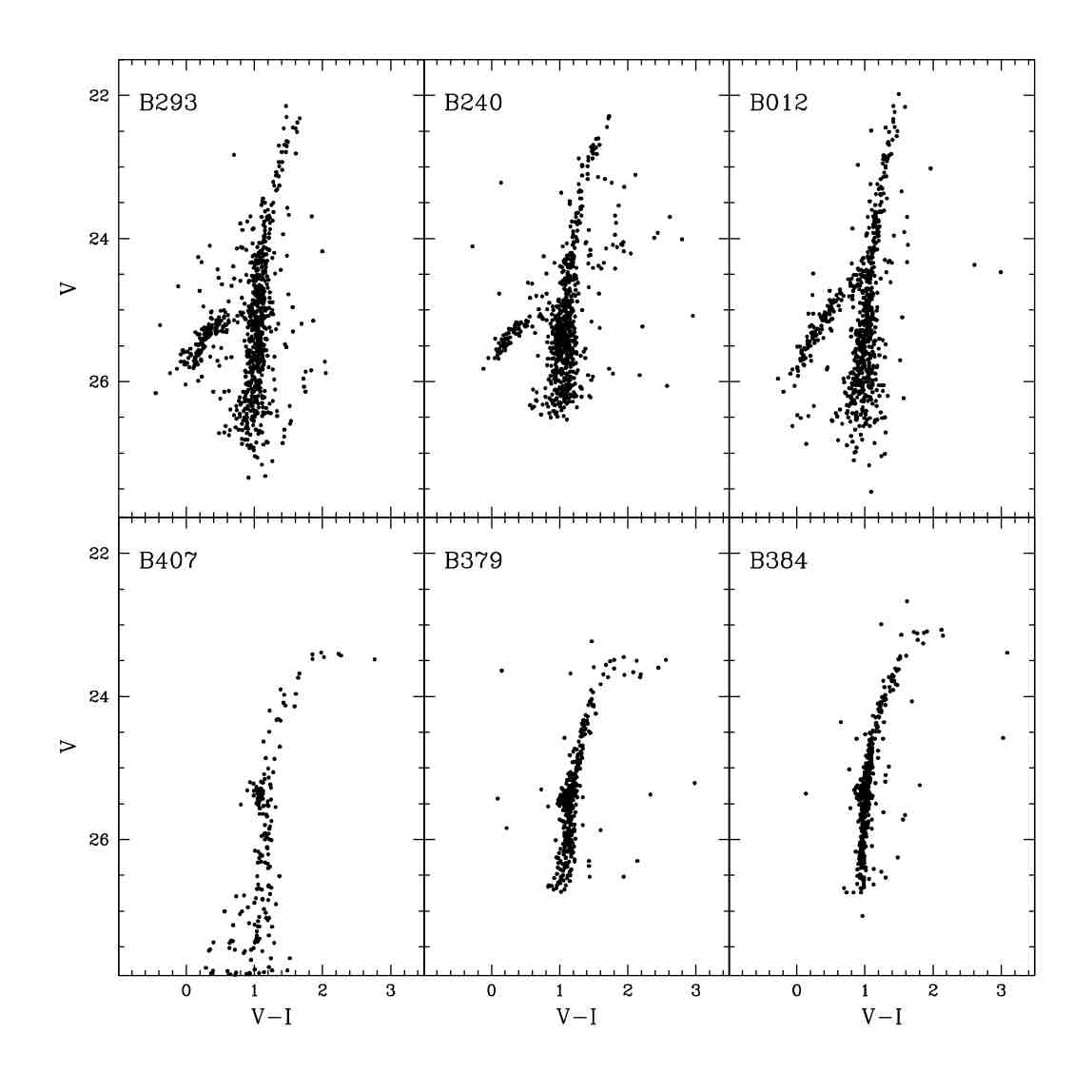}
      \caption{Examples of M31 globular clusters classified as Blue Horizontal Branch (BHB, upper panels) or Red Horizontal Branch (RHB, lower panels). All the CMDs are from 
      Rich et al.~(\cite{rich05}) except B407, which is from P09b.
      }
         \label{cmd6}
   \end{figure}

   \begin{figure}
   \centering
   \includegraphics[width=9cm]{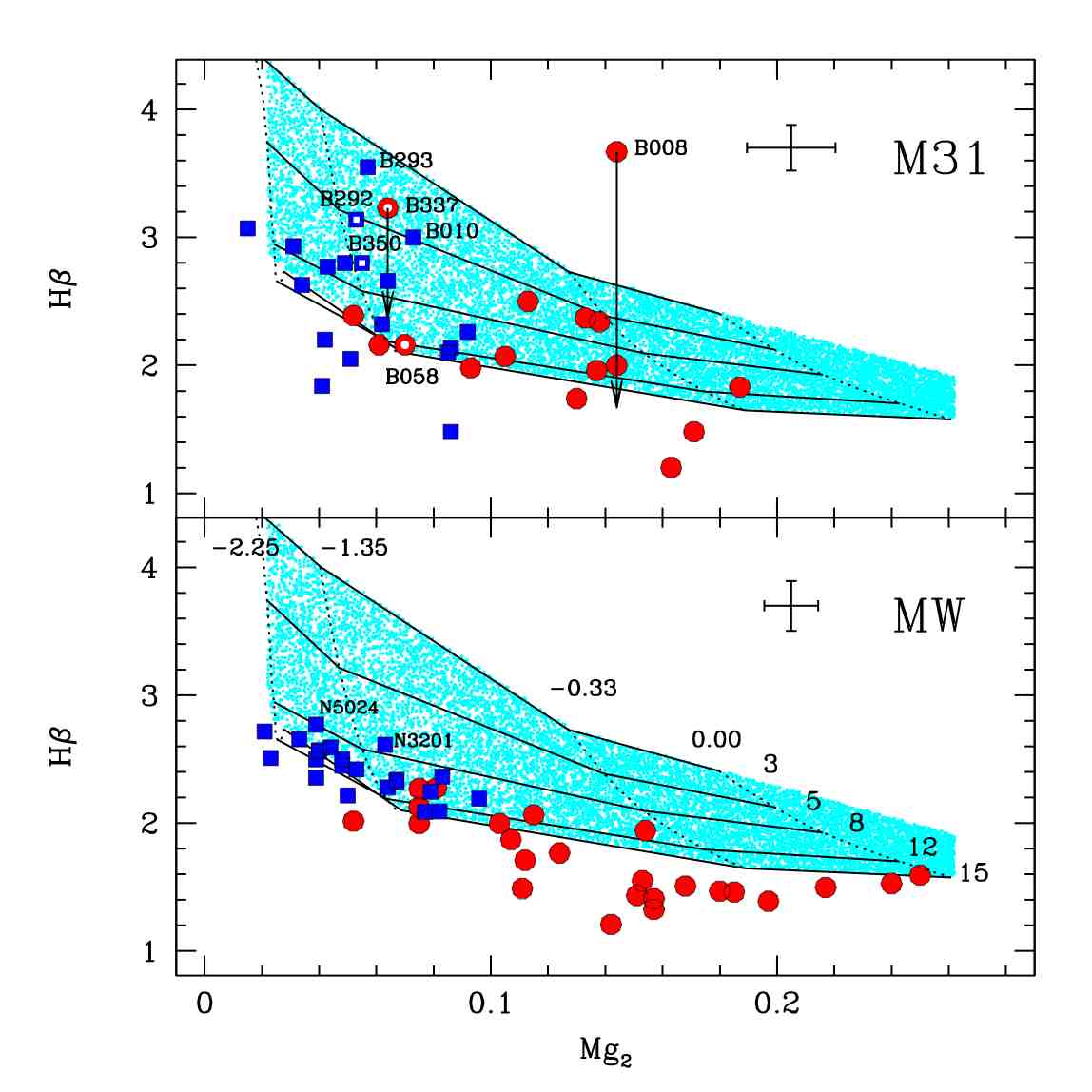}
      \caption{Galactic (lower panel) and M31 (upper panels) globular clusters whose HB morphology can be classified from their CMD in the Mg2 vs. H$\beta$ plane. The average 1~$\sigma$ error-bars are plotted in the upper part of the panels.
      The over-plotted grid of models is from the recent set by Thomas et al.~(\cite{thomas}): continuous lines are iso-age loci of age 3, 5, 8, 12 and 15 Gyr as labeled in the lower panel; dotted lines are iso-metallicity loci
      for [Z/H]=-2.25, -1.35, -0.33, 0.0 and +0.35, as labeled on top of the lower panel. The adopted models are for SSPs with [$\alpha$/Fe]=+0.3. Blue squares are clusters classified as HB-blue, red-circles are clusters classified as HB-red. The main targets of this study are represented with open symbols and labeled. We also labeled three M31 clusters lying above the age=5~Gyr line (B010, B293 and B008) and two Galactic clusters lying on top or above the age=8~Gyr line (NGC5024 and NGC3201). The arrows on the points for B337 and B008 show the position implied by the new estimates of H$\beta$ (Galleti, private communication).
      }
         \label{HbMg2}
   \end{figure}

The sensitivity of colors and/or spectral indices generally used as age diagnostics (like, for example, H$\beta$) to the morphology of the horizontal branch is known for long time: for example, it was already taken into account in the models of the integrated light of SSPs by Buzzoni (\cite{buz}). Bright and hot BHB stars typical of an old population may mimic the effect of a hotter MSTO, i.e., of a younger age (see Lee et al.~\cite{lee} for a thorough discussion, and Percival \& Salaris~\cite{perci} for a very recent investigation in the context of SPSS modeling). It is tempting to identify this as the main reason at the origin of the erroneous classification of B292 and B350 as intermediate-age clusters (Rey et al.~\cite{rey07}).

\begin{figure}
   \centering
   \includegraphics[width=9cm]{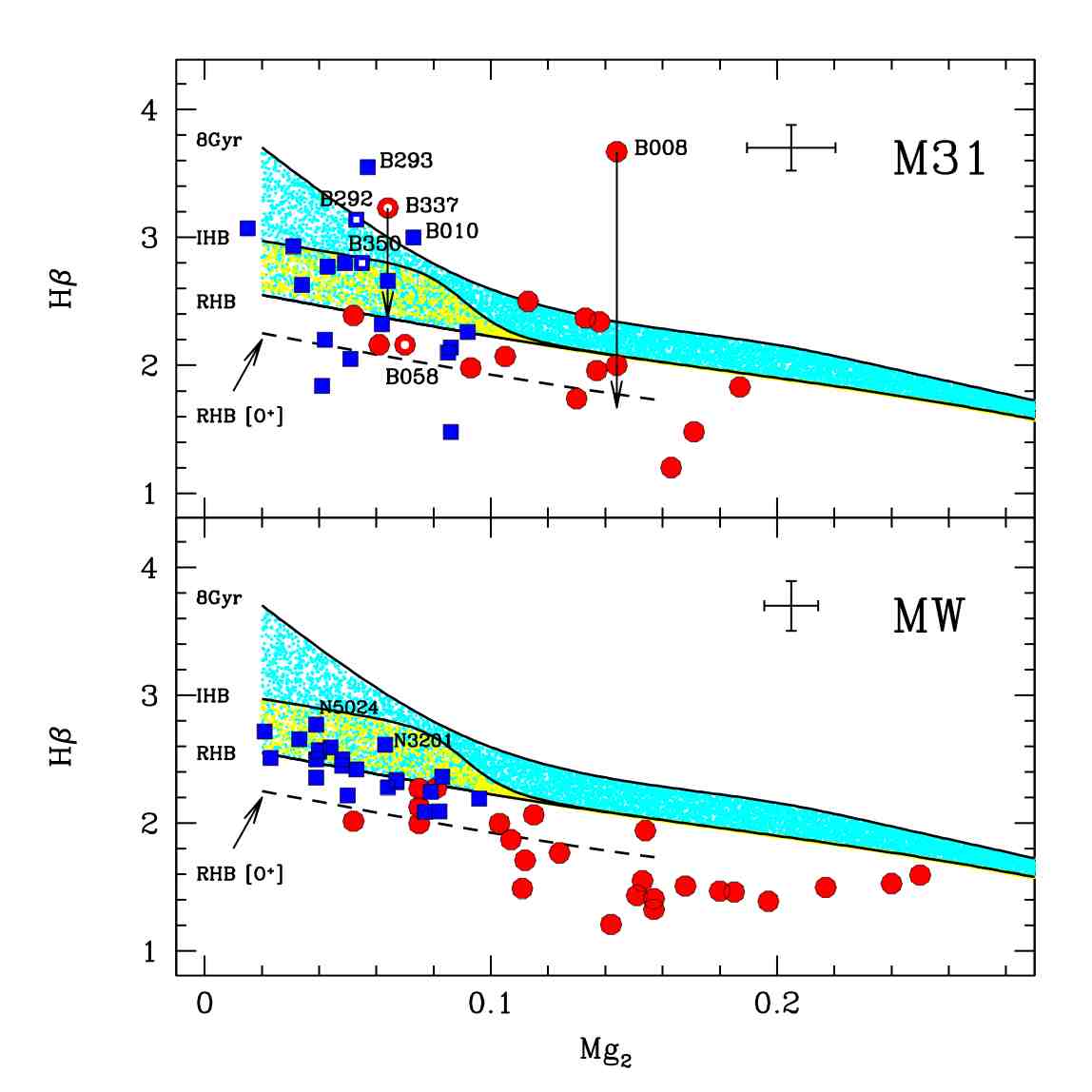}
      \caption{Same as Fig.~\ref{HbMg2} but in this case the comparison is with models from the Buzzoni~(\cite{buz}) set, to illustrate the effects of the HB morphology on H$\beta$.
      The lower continuous line is an age=15~Gyr model that splits in two branches around $Mg_2=0.12$ (i.e., in the low metallicity regime) according to the HB morphology associated to the model, i.e., red (RHB) or intermediate (IHB). The latter was chosen because it provides H$\beta$ values (slightly) higher than the BHB case.  The dashed line is an age=15~Gyr model with enhanced oxygen abundance to show the effect of $\alpha$-enhancement.
      }
         \label{mgbuzz}
   \end{figure}

To gain a deeper insight into this hypothesis, we considered all M31 GCs for which a safe classification of their HB morphology can be obtained from their CMDs, taken from different sources (see Sect.~\ref{int} for an exhaustive list). We simply divided the clusters into BHB (blue squares) and RHB (red circles) from the inspection of their CMD; our classification criterion is illustrated in Fig.~\ref{cmd6}. We ended up with a sample of 33 clusters that also have estimates of the H$\beta$ and Mg2 indices reported in the RBC (see G09). The classical age-diagnostic plot Mg2 vs. H$\beta$ for these clusters is shown in the upper panel of Fig.~\ref{HbMg2}; a grid of SSP models from the most recent set by Thomas et al.~(\cite{thomas}) is over-plotted as a reference. None of the conclusions drawn from this plot is significantly dependent on the actual set of models adopted (see, for example,  Fig.~\ref{mgbuzz}, below). In the lower panel of Fig.~\ref{HbMg2} the same diagram is shown for the Galactic GCs. The H$\beta$ and Mg2 indices are taken from G09. In this case we classified as BHB clusters with $\frac{B-R}{B+R+V}>0.0$ and as RHB those with $\frac{B-R}{B+R+V}<0.0$, where the values of the classical HB morphology parameter\footnote{Where B, R, V are the number of stars to the blue, to the red, and within the RR~Lyrae instability strip, see Lee, Demarque \& Zinn (\cite{leezinn}).} $\frac{B-R}{B+R+V}$ are taken from 
Harris~(\cite{harris}).

In addition to the obvious correlation between the HB morphology and the metallicity of the clusters\footnote{Metallicity is the {\em first} parameter driving the HB morphology, 
see Sandage \& Wildey~(\cite{sand67}), Fusi Pecci \& Bellazzini~(\cite{fpb97}),  
Dotter et al.~(\cite{dotter}), and reference therein.}, there are several other interesting features in Fig.~\ref{HbMg2} that deserve to be commented on in the present context.

\begin{enumerate}

\item According to Fig.~\ref{HbMg2}, B292 and B350 appear younger than 8~Gyr, in agreement with P05. These are not the most extreme cases, because B010 and B293 have even larger H$\beta$ values, implying ages as young as $\sim 3$~Gyr and $\sim 2$~Gyr, respectively, if the comparisons with models shown in Fig.~\ref{HbMg2} are taken at face value. Nevertheless, also these clusters, as B292 and B350 (and also B311, discussed in Sect.~\ref{int}), display an obvious BHB population (see, e.g., Fig.~\ref{cmd6}, for the case of B293) indicating that they are, in fact, much older than this.

\item In this context it is interesting to note that the well studied BHB Galactic clusters NGC5024 and NGC3201 fall, in the same diagram, on top and just above the age=8~Gyr locus, respectively. 
We have reliable age estimates for these clusters based on exquisite photometry that reaches well below the MSTO. In the most recent thorough analysis, based on an homogeneous set of HST data, Dotter et al.~(\cite{dotter}) found an age=12.0$\pm$0.75~Gyr for NGC3201, and an age=13.25$\pm$0.50~Gyr for NGC5024, in an age scale in which the oldest GC has age=13.50$\pm$0.50~Gyr. This shows quite conclusively that even in the best possible case, an age classification based
on the classical spectral diagnostic H$\beta$ is not at all able to unambiguously distinguish between old and intermediate-age clusters, because 13 Gyr old BHB clusters can easily have H$\beta\sim 3\AA$, i.e., the same as $\sim 5-8$ Gyr old clusters (see Lee et al.~\cite{lee}).
In this sense, it appears reasonable the limit H$\beta>3.5\AA$ adopted by Fusi Pecci et al.~(\cite{FP05}, and in the RBC) to identify genuinely {\em young} clusters, even if this a conservative choice may still not be completely safe from false positives, like the case of B008, described below.

\item Returning to M31 BHB clusters, the situation shown in Fig.~\ref{HbMg2} may appear even worse. However, the larger spread of the distribution of M31 points with respect to MW ones is suggestive of the effect of larger observational errors, which would not be unexpected given the large difference in distance between the two sets of GCs. Even if average formal errors provided in the literature are similar in the two cases (see the error-bars in Fig.~\ref{HbMg2}), the fact that the M31 BHB clusters show similar spreads above and below the 12-15 Gyr models is consistent with our hypothesis. 

\item Further support of this view was provided by new spectra that were recently 
obtained for two RHB clusters that appeared to have an anomalously high $H\beta$, B337 (one of our main targets) and B008 (lying straight on the age=1~Gyr locus)\footnote{S. Galleti, private communication. The spectra were obtained at the Cassini Telescope, in Loiano, using the same set-up and data-reduction procedure described in Galleti et al.~(\cite{loi,G09}) The H$\beta$ index was measured as in Galleti et al.~(\cite{G09}). It is found H$\beta=1.67\pm 0.39~\AA$ instead of 
H$\beta=3.67\pm 0.30~\AA$ for B008, and H$\beta=2.37\pm 0.36~\AA$ instead of 
H$\beta=3.23\pm 0.07~\AA$ for B337.}. 
The newly obtained estimates of H$\beta$ for these clusters are much lower than those found in the literature (see the arrows in Fig.~\ref{HbMg2}) and fully agree with ages $\ga 12$~Gyr. This new finding suggest that large errors in Lick indices taken from the literature may not be such rare occurrences for M31 GCs as one might expect.

\end{enumerate}

To illustrate the effects of the HB morphology on H$\beta$, we provide in Fig.~\ref{mgbuzz} the same comparison as in Fig.~\ref{HbMg2} but using models from the set by Buzzoni~(\cite{buz}).
The overall results are similar to those obtained in Fig.~\ref{HbMg2}, hence we do not repeat the discussion. The key point displayed in Fig.~\ref{mgbuzz} is that 
{\em if the observed range of HB morphology is taken into account,
the H$\beta$ values of all Galactic GCs can be reconciled with the old ages estimated from the CMDs} by Dotter et al. (\cite{dotter}).

In conclusion, a convincing case has still to be provided for the presence of an intermediate-age bright cluster in M31 (see Caldwell et al.~\cite{C11}, for a closely similar conclusion, reached with a different observational approach).
The experiment performed here suggests that HST observations of M31 clusters with ``reasonable'' exposure times ($\la 2-3$  HST orbits) can reliably probe only ages lower than 4-5~Gyr, while they would be inconclusive for older ages (except in the case of detection of BHB stars). On the other hand, the results of the present analysis clearly indicate that while the analysis of classical Lick indices remain a valuable technique to identify {\em candidate} intermediate-age GCs in external galaxies (including M31), they do not provide sufficient evidence to firmly establish the case of ages in the range $\sim 5-9$, but need independent confirmation from other means. 

\begin{acknowledgements}
We are grateful to Thomas Puzia for useful discussions.\\
This research is supported by a PRIN-INAF 2009 grant CRA-1.06.12.10 assigned to the 
project ''Formation and Early Evolution of Massive Star Cluster'' (PI: R. Gratton).
\end{acknowledgements}


\begin{thebibliography}{500}

\bibitem[2009]{sdss}
         Abazajian, K.N., Adelman-McCarthy, J.K., Ag\"ueros, M.A., et al., 2009,
	 \apjs, 182, 543

\bibitem[1996]{ajh96}  
         Ajhar,  E.  A.,  Grillmair,  C.  J.,  Lauer,  T.  R.,    Baum,  W.  A., et  al.:  1996,    \aj,    111,  1110

\bibitem[2000]{barm00} 
         Barmby, P., Huchra, J.P., Brodie J.P., Forbes, D.A., 
	 Schroder, L.L., \& Grillmair, C.J, 2000, \aj, 119, 727 [B00]

\bibitem[2001]{barm01} 
         Barmby, P., Huchra, J.P., \& Brodie, J.P. 2001, \aj, 121, 1482

\bibitem[2003]{barm03} 
         Barmby, P., in Extragalactic Globular Cluster Systems, ed. M. Kissler-Patig (Berlin: Springer),
	 2003, 143

\bibitem[2007]{barm07}
        Barmby, P., McLaughlin, D.E., Harris, W.E., 
        \& Harris, G.L.H. \ 2007, \aj, 133, 2764

\bibitem[2002]{beas02} 
         Beasley M. A., Hoyle F., Sharples R. M., 2002, MNRAS, 336,168e 

\bibitem[2005]{beas05} 
         Beasley, M. A., Brodie, J. P., Strader, J., Forbes, D. A., Proctor, R. N., Barmby,
         P., \& Huchra, J. P. 2005, AJ, 129, 1412

\bibitem[2008]{5024}
         Beccari, G., Lanzoni, B., Ferraro, F.R., et al., 2008, \apj, 679, 712

\bibitem[1999]{bel99}
         Bellazzini, M., Ferraro, F. R., \& Buonanno, R. \ 1999, \mnras, 304, 633

\bibitem[2004]{brown04}  
         Brown,  T.M.,  Ferguson,  H.C.,  Smith,  E.,  Kimble,  R.A.,  Sweigart,  A.V., 
         Renzini,  A.,  Rich,  R.M.  \&  VandenBerg,  D.A.  2004b,  \apj, 613, L125            

\bibitem[2003]{bc03}
         Bruzual, G., \& Charlot, S., 2003, \mnras, 344, 1000 (BC03)

\bibitem[1984]{BH84}
         Burstein D., Heiles C. \ 1984, \apjs, 54, 33

\bibitem[2004]{burst04} 
         Burstein, D., et al. 2004, ApJ, 614, 158

\bibitem[1984]{lick1}
         Burstein, D., Faber, S.M., Gaskell, C.M., \& Krumm, N., 1984, \apj,
	 287, 586	 

\bibitem[1989]{buz}	 
         Buzzoni, A. 1989, ApJS, 71, 817

\bibitem[2009]{C09} 
         Caldwell, N., Harding, P., Morrison, H., Rose, J., Schiavon, R., 
	 \& Kriessler, J., 2009, \aj, 137, 94 [C09]
	 
\bibitem[2011]{C11}
         Caldwell, N., Schiavon, R., Morrison, H., Rose, J.A., Harding, 2011, \aj, 141, 61	
	  
\bibitem[2001]{clem01} 
         Clementini, G., Federici, L., Corsi, C., Cacciari, C., Bellazzini, M., 
	 \& Smith, H.A., 2001, \aj, 559, L109  

\bibitem[2000a]{dol_a} 
         Dolphin, A. E. 2000a, PASP, 112, 1383

\bibitem[2000b]{dol_b} 
         Dolphin, A. E. 2000b, PASP, 112, 1397

\bibitem[2010]{dotter} 
         Dotter, A., Sarajedini, A., Anderson, J., et al., 2010, \apj, 708, 698

\bibitem[1985]{lick2}
         Faber, S. M., Friel, E. D., Burstein, D., \& Gaskell, C. M., 1985, ApJS, 57, 711

\bibitem[2006]{fan06} 
         Fan, Z., Ma, J., de Grijs, R., Yang, Y., \& Zhou, X. 2006,
                      MNRAS, 371, 1648

\bibitem[2008]{fan08} 
         Fan, Z., Ma, J., de Grijs1, R., \& Zhou, X., 2008, 
	 MNRAS, 385, 1973

\bibitem[2010]{fan10} 
         Fan, Z., de Grijs1, R., \& Zhou, X., 2010, \mnras, in press 
	 (arXiv:1009.3582) [F10]

\bibitem[1993]{fp93}
         Fusi Pecci, F., Ferraro, F.R., Bellazzini, M., Djorgovski, S., Piotto,
	 G., \& Buonanno, R., 1993, \aj, 105, 1145

\bibitem[1997]{fpb97}
         Fusi Pecci, F., \& Bellazzini, M., 1997, in The Third Conference on 
	 Faint Blue Stars (1997), A.G.D. Philip, J. Liebert, R. Saffer and 
	 D.S. Hayes Eds., Shenectady: L. Davis Press, p. 255

\bibitem[1996]{fp96} 
         Fusi Pecci, F., et al. 1996, AJ, 112, 1461 
	 
\bibitem[2005]{FP05} 
         Fusi Pecci, F., Bellazzini, M., Buzzoni, A., De Simone, E., Federici, L., \&
         Galleti, S. 2005, \aj, 130, 554
	 
\bibitem[2002]{isocPD} 
         Girardi, L., Bertelli, G., Bressan, A., Chiosi, C., 
         Groenewegen, M.A.T., Marigo, P., Salasnich, B., 
         \& Weiss, A. 2002, \aap, 391, 195
	 
\bibitem[2004]{rbc}
         Galleti, S., Federici, L., Bellazzini, M., et al., 2004, \aap, 416, 917

\bibitem[2005]{loi}
         Galleti, S., Bellazzini, M., Federici, L., \& Fusi Pecci, 2005, 
	 A\&A, 436, 535

\bibitem[2006]{b514} 
         Galleti, S., Federici, L., Bellazzini, M., Buzzoni, A., 
	 Fusi Pecci, F. 2006, \apj, 650, 107

\bibitem[2009]{G09} 
         Galleti, S., Bellazzini, M., Buzzoni, A., Federici, L., Fusi Pecci, F. 2009, \aap, 508, 1285 [G09]    

\bibitem[1996]{harris}
         Harris, W.E., 1996, \aj, 112, 1487

\bibitem[1997]{holl97} 
         Holland, S., Fahlman, G. G., \& Richer, R. B. 1997, \aj, 114, 1488 (HFR) 

\bibitem[1991]{huch91} 
         Huchra, J. P., Kent, S. M., \& Brodie, J. P. 1991, \apj, 370, 495

\bibitem[2004]{hux04} 
         Huxor, A., Tanvir, N. R., Irwin, M., Ferguson, A., Ibata, R., 
	 Lewis, G., Bridges, T. 2004, ASPC, 327, 118

\bibitem[2005]{hux05} 
         Huxor, A. P., Tanvir, N. R., Irwin, M. J., Ibata, R., Collett, J. L., 
	 Ferguson, A. M. N., Bridges, T., Lewis, G. F. 2005, \mnras, 360, 1007

\bibitem[2008]{hux08}
         Huxor, A. P., Tanvir, N. R., Ferguson, A. M. N., 
         Irwin, M. J., Ibata, R., Bridges, T., Lewis, G. F. 2008, \mnras, 385, 1989

\bibitem[2001]{ibata01} 
         Ibata, R., Irwin,M., Lewis, G., Ferguson, A. M. N., \& Tanvir, N. 
	 2001a, Nature, 412, 49 

\bibitem[2000]{jab00}  
         Jablonka,  P.,  Courbin,  F.,  Meylan,  G.,  Sarajedini,  A.,  
	 Bridges,  T.J. \&  Magain,  P.    2000,  \aap,  359,  131 

\bibitem[2003]{jiang03} 
         Jiang, L., Ma, J., Zhou, X., Chen, J., Wu, H., \& Jiang, Z. 
	 2003, \aj, 125, 727

\bibitem[2009]{3201}
         Kravtsov, V., Alca\'ino, G., Marconi, G., Alvarado, F., 2009, \aap,
	 497, 371

\bibitem[1990]{leezinn}
         Lee, Y.-W., Demarque, P., Zinn, R.J., 1990, \apj, 350, 155	 

\bibitem[2000]{lee}
         Lee, H.-c., Yoon, S.-Y., Lee, Y.-W., 2000, \aj, 120, 998

\bibitem[2007]{ma07} 
         Ma, J., et al. 2007, \apj, 659, 359

\bibitem[2009]{ma09} 
         Ma, J., et al. 2009, \aj, 137, 4884
 
\bibitem[2010]{ma10}
         Ma, J., et al., 2010, \pasp, 122, 1164

\bibitem[1998]{mara98} 
         Maraston, C. 1998, MNRAS, 300, 872
	 
\bibitem[2005]{mara05} 
         Maraston, C. 2005, MNRAS, 362, 799

\bibitem[2005]{mcc05}
         McConnachie, A.W., Irwin, M.J. \& Ferguson, A.M.N. \ 2005, \mnras, 356, 979

\bibitem[2006]{mack06} 
         Mackey, A. D., Huxor, A., Ferguson, A. M. N., Tanvir, N. R., Irwin, M., Ibata, R., Bridges, T., Johnson, R. A., Lewis, G. 2006, \apj, 653, 105

\bibitem[2007]{mack07} 
         Mackey, A. D., Huxor, A., Ferguson, A. M. N., Tanvir, N. R., Irwin, M., Ibata, R., Bridges, T., Johnson, R. A., Lewis, G. 2007, \apj, 655, 85

\bibitem[2010]{mack10}
         Mackey, A.D., Huxor, A.P., Ferguson, A.M.N., et al., 2010, \apj, 717,
	 L11

\bibitem[2006]{martin06} 
         Martin, N. F., Ibata, R. A., Irwin, M. J., Chapman, S., Lewis, G. F., Ferguson, A. M. N., Tanvir, N., \& McConnachie, A. W. 2006, \mnras, 371, 1983 

\bibitem[2001]{mey01} 
         Meylan, G., Sarajedini, A., Jablonka, P., Djorgovski, S. G., Bridges, T., \& Rich, R. M. 2001, \aj, 122, 830 


\bibitem[1988]{rfp}
         Renzini, A., \& Fusi Pecci, F. 1988, \araa, 26, 199

\bibitem[2007]{rey07} 
         Rey, S.-C. et al. 2007, ApJS, 173, 643
	 
\bibitem[1996]{ric96a}  
         Rich,  R.M.,  Mighell,  K.J.,  Freedman,  W.,  \&  Neill,  J.D.  1996, \aj,  111,  768 

\bibitem[2003]{rich03} 
         Rich, R.M., in ASP Conf. Ser. 296, New Horizons in Globular Cluster Astronomy, ed. G.     Piotto et al. (San Francisco: ASP), 
	 2003, 533

\bibitem[2005]{rich05}
         Rich, R.M, Corsi, C.E., Cacciari, C., Federici, L., Fusi Pecci,
	 Djorgovski, S.G., \& Freedman, W., 2005, \aj, 129, 2670 
	 
\bibitem[2010]{perci}
         Percival, S.M., Salaris, M., 2010, \mnras, in press (arXiv:1012.0004)	  
	 
\bibitem[2009a]{P09a} 
         Perina, S., Barmby, P., Beasley, M.A., et al., 2009a, \aap, 494, 933	 

\bibitem[2009b]{P09b} 
         Perina, S., Federici, L., Bellazzini, M., Fusi Pecci, F., Cacciari, C.,
	 \& Galleti, S., 2009b, \aap, 507, 1375 [P09b]

\bibitem[2010]{ymc} 
         Perina, S., Bellazzini, M., Barmby, P., Cohen, J. G., Hodge, P. W.,
	 Puzia, T. H., \& Strader, J. 2010, \aap, 511, 23

\bibitem[2002]{pio02}
         Piotto, G., King, I.R., Djorgovski, S.G., Sosin, C., Zoccali, M. et al. 
          \ 2002,  \aap, 391, 945

\bibitem[2002]{puzia02} 
         Puzia, T.H, Saglia, R.P., Kissler-Patig, M., Maraston, C., Greggio, L., Renzini, A., \& Ortolani, S. 
	 2002, \aap, 395, 45

\bibitem[2005]{puzia} 
         Puzia, T.H, Perrett, K.M., Bridges, T.J., 
	 2005, \aap, 434, 909 [P05]

\bibitem[1967]{sand67}
         Sandage, A., \& Widley, R., 1967, \apj, 150, 469  


\bibitem[2007]{saraj07} 
         Sarajedini A., Mancone C.L., \aj, 134, 447 

\bibitem[1998]{sch98} 
         Schlegel, D.J., Finkbeiner, D.P., \&  Davis, M.  1998, \apj, 500, 525 

\bibitem[2005]{siria} 
         Sirianni, M., Jee, M. J., Benitez, N., et al. 2005, PASP, 117, 1049

\bibitem[2009]{strad09} 
         Strader, J., Smith, G. H., Larsen, S., Brodie, J. P., Huchra, 
         J.P., 2009, \aj, 138, 547

\bibitem[2003]{thomas03}
         Thomas, D., Maraston, C., Bender, R., 2003, \mnras, 339, 897 (T03)

\bibitem[2010]{thomas} 
         Thomas, D., Maraston, C., Johansson, J., 2010, \mnras, in press (arXiv:1010.4569)

\bibitem[2010]{wang10} 
         Wang, S., Fan, Z., Ma, J., de Grijs R., Zhou, X., 
	 2010, \aj, 139, 1438 [W10]

\bibitem[1994]{worthey}
         Worthey, G., 1994, ApJS, 95, 107

\end{thebibliography}
\end{document}